\documentclass[traditabstract]{aa}
\usepackage{graphicx}
\usepackage{txfonts}
\usepackage{natbib}

\renewcommand{\d}{\mathrm{d}}
\newcommand{\e}{\mathrm{e}}
\newcommand{\be}{\begin{equation}}
\newcommand{\ee}{\end{equation}}

\begin{document} 

\title{An analytic approach to number counts of weak-lensing peak detections}

\author{Matteo Maturi\thanks{\email{maturi@ita.uni-heidelberg.de}}
  \and Christian Angrick
  \and Matthias Bartelmann
  \and Francesco Pace}

\titlerunning{Weak-lensing peak detections}
\authorrunning{Maturi et al.}

\institute{Zentrum f\"ur Astronomie der Universit\"at Heidelberg, Institut f\"ur Theoretische Astrophysik, Albert-Ueberle-Str.~2, 69120 Heidelberg, Germany}

\date{\emph{Astronomy \& Astrophysics, submitted}}

\abstract{We develop and apply an analytic method to predict peak
  counts in weak-lensing surveys. It is based on the theory of
  Gaussian random fields and suitable to quantify the level of
  spurious detections caused by chance projections of large-scale
  structures as well as the shape and shot noise contributed by the
  background galaxies. We compare our method to peak counts obtained
  from numerical ray-tracing simulations and find good agreement at
  the expected level. The number of peak detections depends
  substantially on the shape and size of the filter applied to the
  gravitational shear field. Our main results are that weak-lensing
  peak counts are dominated by spurious detections up to
  signal-to-noise ratios of 3--5 and that most filters yield only a
  few detections per square degree above this level, while a filter
  optimised for suppressing large-scale structure noise returns up to
  an order of magnitude more.
  \keywords{Cosmology: theory -- large-scale structure of Universe --
    Galaxies: clusters: general -- Gravitational lensing} }

\maketitle

\section{Introduction}

Wide-area surveys for weak gravitational lensing can be and have been
used for counting peaks in the shear signal, which are commonly
interpreted as the signature of sufficiently massive dark-matter
halos. However, such detections are clearly contaminated by spurious
detections caused by the chance superposition of large-scale
structures, and even by the shape- and shot-noise contributions from
the background galaxies used to sample the foreground shear field. As
a function of the peak height, what is the contribution of genuine
halos to these detections, and how much do the large-scale structure
and the other sources of noise contribute?

Given the power of lensing-peak number counts as a cosmological probe
\citep{MA09.1,KR09.1,DI09.1}, we address this question here after
developing a suitable analytic approach based on peak counts in
Gaussian random fields, as laid out by \cite{BA86.1}. It is reasonable
to do so even though at least the high peaks are caused by halos in
the non-Gaussian tail of the density fluctuations because the noise
and large-scale structures contributions remain Gaussian, and thus at
least the contamination of the counts can be well described
analytically. Peaks with the highest signal-to-noise ratios are
expected to be more abundant than predicted based on Gaussian random
fields.

Weak-lensing data are filtered to derive peak counts from
them. Several linear filters have been proposed and used in the
literature. They can all be seen as convolutions of the measured shear
field with filter functions of different shapes. Many shapes have been
proposed for different purposes \citep{SC98.2,SC04.2,MAT04.2}. One
filter function, called the optimal filter later on, was designed
specifically to suppress the contribution from the large-scale
structure by maximising the signal-to-noise ratio of halo detections
against the shear field of the large-scale structure.

We study three such filters here, with the optimal filter among
them. Results will differ substantially, arguing for a careful filter
choice if halo detections are the main goal of the application. We
compare our analytic results to a numerical simulation and show that
both agree at the expected level. We begin in \S~2 with a brief
summary of gravitational lensing as needed here and describe filtering
methods in \S~3. We present our analytic method in \S~4 and compare it
to numerical simulations in \S~5, where we also show our main
results. Conclusions are summarised in \S~6. In Appendix~A, we show
predictions of peak counts and the noise levels in them for several
planned and ongoing weak-lensing surveys.

\section{Gravitational lensing}

Isolated lenses are characterised by their \textit{lensing potential}
\begin{equation}\label{eq:l_potential}
  \psi(\vec{\theta}) \equiv \frac{2}{c^2}
                            \frac{D_{\rm ds}}{D_{\rm d}D_{\rm s}}
                            \int 
			    \Phi(D_{\rm d}\vec{\theta}, z)\,\d z \;,
\end{equation}
where $\Phi$ is the Newtonian gravitational potential and $D_{\rm s,
  d, ds}$ are the angular-diameter distances between the observer and
the source, the observer and the lens, and the lens and the source,
respectively.
The potential $\psi$ relates the angular positions $\vec\beta$ of the
source and $\vec\theta$ of its image on the observer's sky through the
\textit{lens equation}
\begin{equation}\label{eq:lensequation}
  \vec{\beta}=\vec{\theta}-\vec\nabla\psi\;.
\end{equation}
Since sources such as distant background galaxies are much smaller
than the typical scale on which the lens properties change and the
angles involved are small, it is possible to linearise
Eq.~({\ref{eq:lensequation}}) such that the induced image distortion
is expressed by the Jacobian
\begin{equation}
  A = (1-\kappa)
      \left(
            \begin{array}{cc}
	      1-g_1 & g_2 \\
	      g_2 & 1+g_1 \\
	    \end{array}
      \right)\,,
\end{equation} 
where $\kappa=\nabla^2 \psi/2$ is the \textit{convergence} responsible
for the isotropic magnification of an image relative to its source,
and $g(\vec{\theta})=\gamma(\vec{\theta})/[1-\kappa(\vec{\theta})]$ is
the \textit{reduced shear} quantifying its distortion. Here,
$\gamma_1=\left(\psi_{,11}-\psi_{,22}\right)/2$ and
$\gamma_2=\psi_{,12}$ are the two components of the \textit{complex
  shear}. Since the angular size of the source is unknown, only the
reduced shear can be estimated starting from the observed ellipticity
of the background sources,
\begin{equation}
  \epsilon=\frac{\epsilon_{\rm s}+g}{1+g^*\epsilon_{\rm s}}\;,
  \label{eq:epstr}
\end{equation}
where $\epsilon_\mathrm{s}$ is the intrinsic ellipticity of the source
and the asterisk denotes complex conjugation.

\section{Measuring weak gravitational lensing}

\subsection{Weak lensing estimator}

In absence of intrinsic alignments between background galaxies due to
possible tidal interactions \citep{HE88.1,SC09.1}, the intrinsic
source ellipticities in Eq.~(\ref{eq:epstr}) average to zero in a
sufficiently large source sample.
An appropriate and convenient measure for the lensing signal is the
weighted average over the tangential component of the shear
$\gamma_{\rm t}$ relative to the position $\vec{\theta}$ on the sky,
\begin{equation}
  \tilde A(\vec\theta)=\int\d^2\theta'\gamma_{\rm t}(\vec{\theta'},\vec{\theta})
  Q(|\vec{\theta'}-\vec{\theta}|)\;.
  \label{eq:A}
\end{equation}
The filter function $Q$ determines the statistical properties of
the estimator $\tilde A$. We shall consider three filter functions
here which will be described in \S~\ref{sec:filters}.

Data on gravitational lensing by a mass concentration can be modeled
by a signal $s(\vec \theta)=A\tau(\vec{\theta})$ described by its
amplitude $A$ and its radial profile $\tau$, and a noise component
$n(\vec{\theta})$ with zero mean, e.g.
\begin{equation}
  \gamma_{\rm t}(\vec{\theta})=A\tau(\vec{\theta})+n(\vec{\theta})
\end{equation}
for the tangential shear.
The variance of the estimator $\tilde A$ in (\ref{eq:A}) is
\begin{equation}
  \sigma^2_{\tilde A}=\int\frac{k\,\mathrm{d} k}{2\pi}P(k)
  \hat W^2(k)|\hat Q(\vec{k})|^2\;,
  \label{eq:sigma_A}
\end{equation}
where $\hat W(k)$ is the \textit{frequency response} of the survey
depending on its geometrical properties, $\hat Q(\vec{k})$ is the
Fourier transform of the filter $Q$, and $P(k)$ is the power spectrum
of the noise component.

\subsection{Weak lensing filters}\label{sec:filters}

\begin{figure*}[!ht]
  \centering
  \includegraphics[angle=-90,width=0.32\hsize]{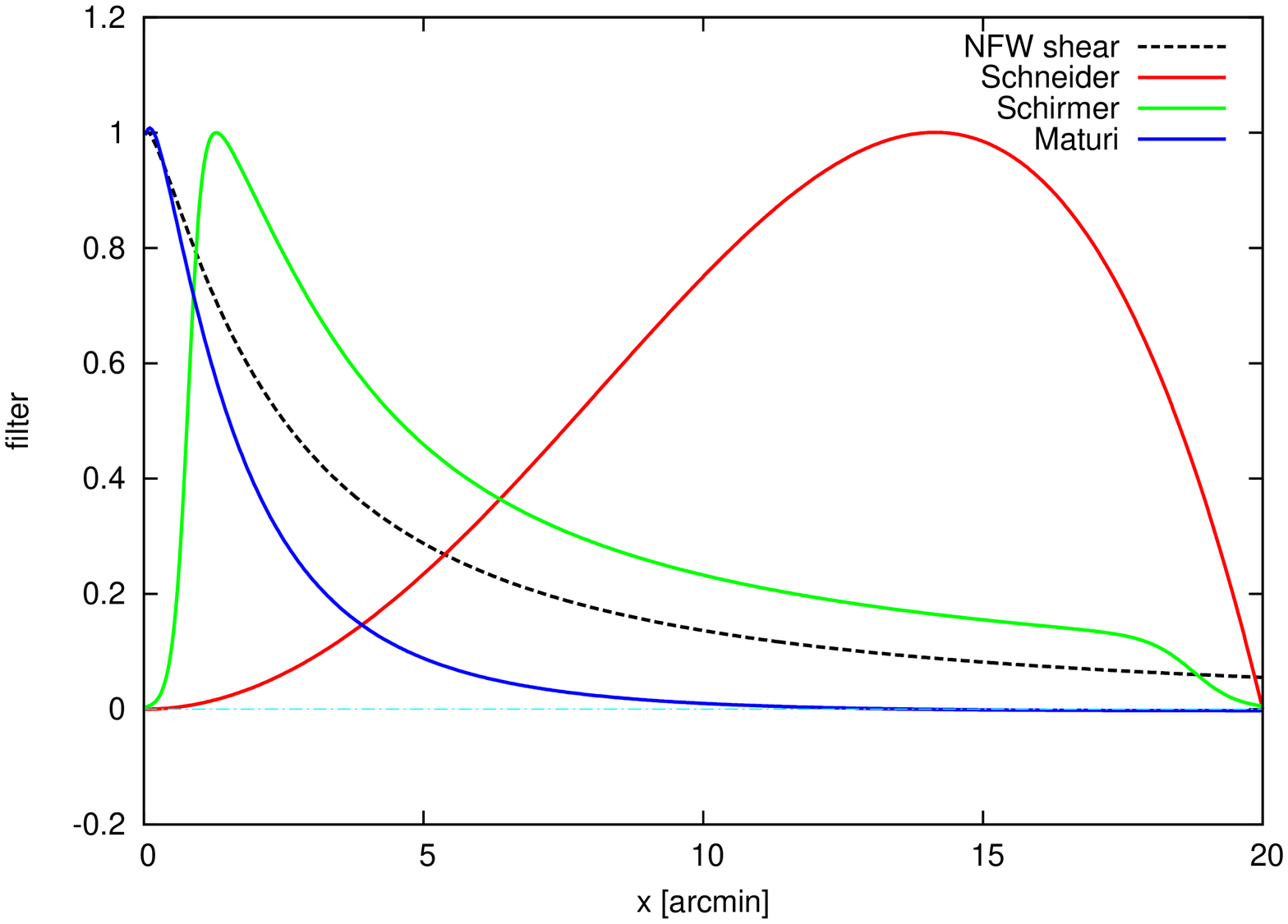}
  \includegraphics[angle=-90,width=0.32\hsize]{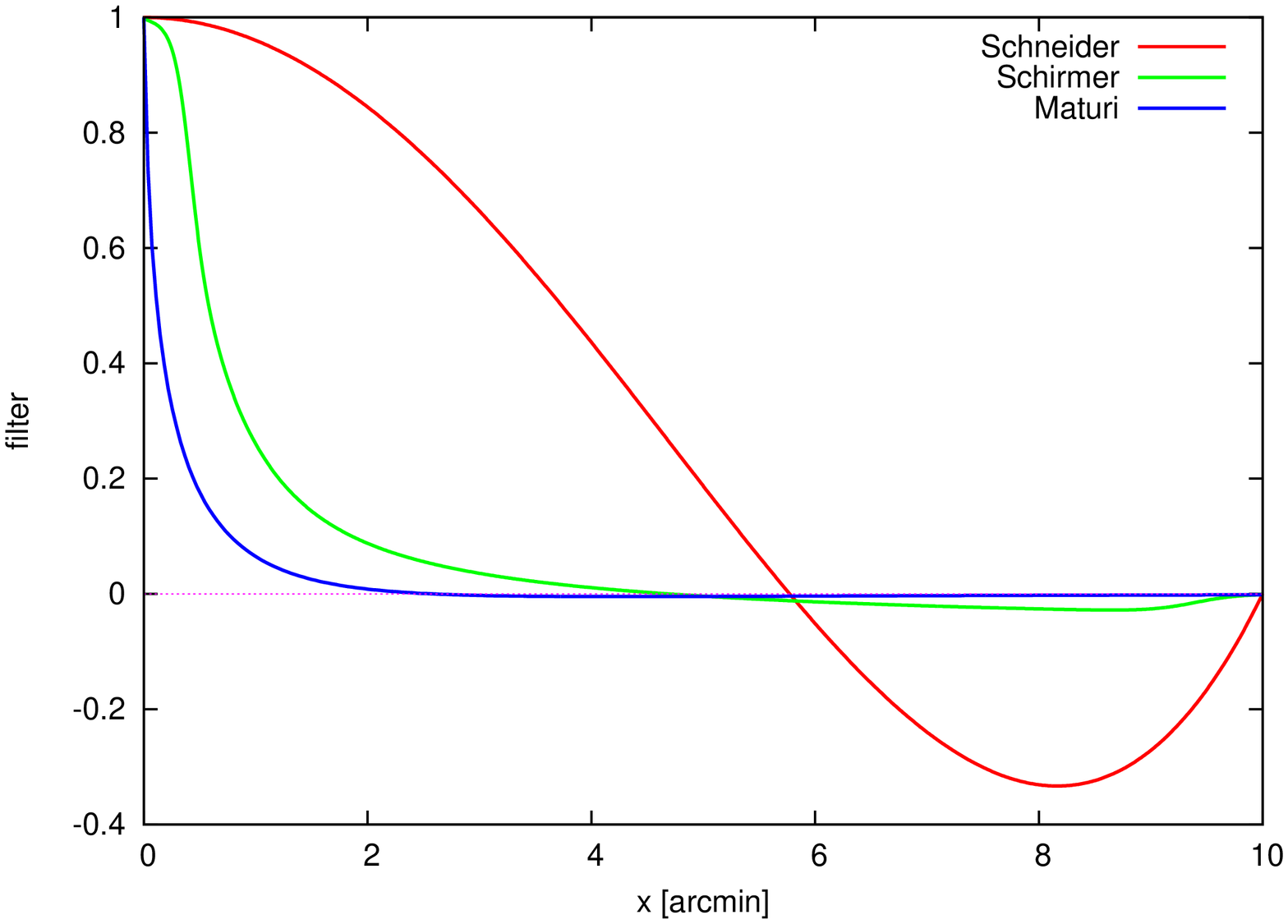}
  \includegraphics[angle=-90,width=0.32\hsize]{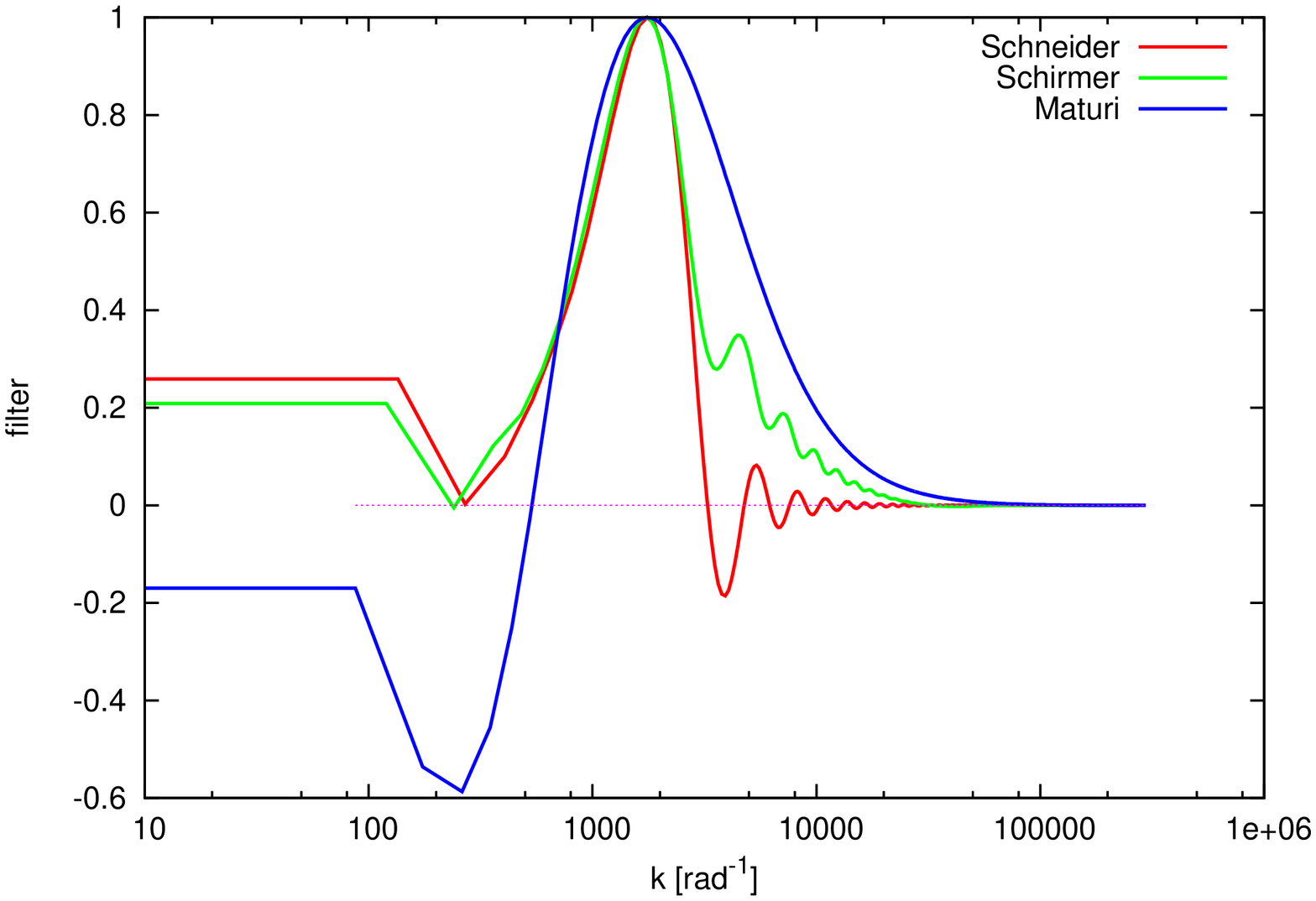}
  \caption{Overview of different weak-lensing filters. The left panel
    shows the three filters adopted here to be used on shear
    catalogues, while the central and right panels show the
    corresponding filters to be used on convergence fields both in
    real and Fourier space, respectively. For illustration only, the
    spatial frequencies are rescaled such that the main filters peaks
    coincide.}
\label{fig:filters}
\end{figure*}

Different filter profiles have been proposed in the literature
depending on their specific application in weak-lensing. We adopt
three of them here which have been used so far to identify halo
candidates through weak lensing.

(1) The polynomial filter described by \cite{SC98.2}
\begin{equation}
  Q_{\rm poly}(x)=\frac{6x^2}{\pi r_{\rm s}^2}\left(1-x^2\right)
  H\left(1-x\right)\;,
\end{equation}
where the projected angular distance from the filter centre,
$x=r/r_{\rm s}$, is expressed in units of the filter scale radius,
$r_{\rm s}$, and $H$ is the Heaviside step function. This filter was
originally proposed for cosmic-shear analysis but several authors have
used it also for dark matter halo searches \citep[see for
e.g.][]{ER00.1,SC04.2}.

(2) A filter optimised for halos with NFW density profile,
approximating their shear signal with a hyperbolic tangent
\citep{SC04.2},
\begin{equation}
  Q_{\rm tanh}(x)=\left(1+\e^{a-bx}+\e^{cx-d}\right)^{-1}
  \tanh\left(x/x_{\rm c}\right)\;,
  \label{eq:mischas}
\end{equation}
where the two exponentials in parentheses are cut-offs imposed at
small and large radii ($a=6$, $b=150$, $c=50$, and $d=47$) and $x_{\rm
  c}$ is a parameter defining the filter-profile slope. A good
choice for the latter is $x_{\rm c}=0.1$ as empirically shown by
\cite{HE05.2}.

(3) The optimal linear filter introduced by \cite{MAT04.2} which,
together with the optimisation with respect to the expected
halo-lensing signal, optimally suppresses the contamination due to the
line-of-sight projection of large-scale structures (LSS),
\begin{equation}  \label{eq:optimal}
  \hat Q_{\rm opt}(\vec k) =
  \alpha
  \frac{\tau(\vec{k})}{P(k)}
  \quad\mbox{with}\quad
  \alpha^{-1}=
  \int \d^2 k \frac{\left|\hat\tau (\vec{k})\right|^2}{P(k)}\;.
\end{equation}
Here, $\hat\tau(k)$ is the Fourier transform of the expected shear
profile of the halo and $P(k)=P_{\rm g}+P_{\rm LSS}(k)$ is the
complete noise power spectrum including the LSS through $P_{\rm LSS}$
and the noise contributions from the intrinsic source ellipticities
and the shot noise by $P_{\rm g}=\sigma_{\rm g}^2/(2n_{\rm g})$, given
their angular number density $n_{\rm g}$ and the intrinsic ellipticity
dispersion $\sigma_{\rm g}$.
This filter depends on parameters determined by physical quantities
such as the halo mass and redshift, the galaxy number density and the
intrinsic ellipticity dispersion and not on an arbitrarily chosen
scale which has to be determined empirically through costly numerical
simulations \citep[e.g.][]{HE05.1}.
An application of this filter to the GaBoDS survey \citep {SC03.2} was
presented in \cite{MAT05.3}, while a detailed comparison of these
filters was performed by \cite{PA07.1} by means of numerical
ray-tracing simulations. They found that the optimal linear filter
given by Eq.~(\ref{eq:optimal}) returns the halo sample with the
largest completeness ($100\%$ for masses $M\geq 3\times 10^{14}\,
  h^{-1}\,M_\odot$ and $\sim 50\%$ for masses $M\sim 2\times 10^{14}\,
  h^{-1}\,M_\odot$ for sources at $z_\mathrm{s}=1$) and the smallest number of
  spurious detections caused by the LSS ($\leq 10\%$ for a
  signal-to-noise threshold of $\mathrm{S/N}\sim 5$).

\subsection{Weak lensing estimator and convergence}

In order to simplify comparisons with numerical simulations, we
convert the estimator $\tilde A$ from Eq.~(\ref{eq:A}) to an estimator
of the convergence,
\begin{eqnarray} \label{eq:A_U}
  \tilde A(\vec\theta) & = & \int\d^2\theta'\kappa(\vec\theta')
  U(|\vec\theta'-\vec\theta|)\;,
\end{eqnarray}
where $U$ is related to $Q$ by
\begin{equation}\label{eq:psi-u}
  Q(\theta) =
  \frac{2}{\theta^2}\int_0^{\theta}\d\theta'\theta'U(\theta')-U(\theta)
\end{equation}
\citep{SC96.2} if the weight function $U(\theta)$ is defined to be
compensated, i.e.
\begin{equation}
\label{eq:comp}
  \int\d\theta'\theta'U(\theta')=0\;.
\end{equation}
Equation~(\ref{eq:psi-u}) has the form of a Volterra integral equation
of the first kind which can be solved with respect to $U$ once $Q$ is
specified. If $\lim_{\rm x\rightarrow 0} Q(x)/x$ is finite, the
solution is
\begin{equation}\label{eq:Psi2U}
  U(\theta)=-Q(\theta)-\int_0^\theta \d\theta^\prime
  \frac{2}{\theta^\prime}Q(\theta^\prime) \;,
\end{equation}
\citep{PO98.1} which can be solved analytically for the polynomial filter
\begin{equation}
  U_{\rm poly}(x)=\frac{9}{\pi r_{\rm s}^2}\left(1-x^2\right)
  \left(\frac{1}{3}-x^2\right)
  H\left(1-x\right)\;,
\end{equation}
and numerically for the hyperbolic-tangent filter of
Eq.~(\ref{eq:mischas}) with an efficient recursive scheme over the
desired radii $\theta$.
If $\lim_{\rm x\rightarrow 0} Q(x)/x=\infty$ as in the case of the
optimal filter, Eq.~(\ref{eq:Psi2U}) can be solved by introducing an
exponential cut-off at small radii to avoid the divergence. The
correct solution is obtained if the cut-off scale is close to the mean
separation between the background galaxies, so that no information is
lost. Alternatively, Eq.~(\ref{eq:psi-u}) can be solved iteratively
with respect to $Q$ by
\begin{equation}
  U_0(\theta)=-Q(\theta) \;,
  \quad
  U_n(\theta)=-Q(\theta)+\frac{2}{\theta^2}\int_0^\theta \d\theta'\theta' U_{n-1}(\theta) \;.
  \label{eq:ebase}
\end{equation}
The iterative procedure is stopped once the difference
$U_n(\theta)-U_{n-1}(\theta)$ is sufficiently small. After $U(\theta)$
has been found, an appropriate constant $c$ has to be added in order
to satisfy the compensation requirement, Eq.~(\ref{eq:comp}). It is
given by
\begin{equation}
\label{eq:constant}
  c=-\frac{2}{\theta_\mathrm{max}^2}\int\limits_0^{\theta_\mathrm{max}}\d\theta^\prime
  \theta^\prime U(\theta^\prime)\;.
\end{equation}

We show in Fig.~(\ref{fig:filters}) the resulting filter profiles to
be used on shear catalogues through Eq.~(\ref{eq:A}) and their
corresponding variants to be used on convergence fields with
Eq.~(\ref{eq:A_U}) both in real and Fourier space. All of them are
band-pass filters and the two of them designed for halo searches have
larger amplitudes at higher frequencies compared to the polynomial
filter by \cite{SC98.2}, where the halo signal is most
significant. This feature is particularly prominent for the optimal
filter, which is additionally negative at low frequencies, where the
LSS signal dominates. These two features ensure the minimisation of
the LSS contamination in halo searches.

\section{Predicting weak lensing peak counts}\label{sec:prediction}

Our analytic predictions for the number counts of weak-lensing
detections as a function of their signal-to-noise ratio are based on
modelling the analysed lensing data, resulting from
Eq.~(\ref{eq:A_U}), as an isotropic and homogeneous Gaussian random
field.
This is an extremely good approximation for the noise and the LSS
components, but not for the non-linear structures such as sufficiently
massive halos, as we shall discuss in Sec.~\ref{sec:comparison}.

\subsection{Statistics of Gaussian random fields}
\label{sec:randomf}

An \emph{$n$-dimensional random field} $F(\vec{r})$ assigns a set of
random numbers to each point $\vec r$ in an $n$-dimensional space. A
joint probability function can be declared for $m$ arbitrary points
$\vec{r}_j$ as the probability to have field values between
$F(\vec{r}_j)$ and $F(\vec{r}_j)+{\rm d}F(\vec{r}_j)$, with
$j=1,\ldots,m$. For \emph{Gaussian random fields}, the field itself,
its derivatives, integrals and any linear combination thereof are
Gaussian random variables which we denote by $y_i$ with mean value
$\langle y_i\rangle$ and central deviations $\Delta y_i:=y_i-\langle
y_i\rangle$, with $i=1,\ldots,p$. Their joint probability function is
a multivariate Gaussian,
\begin{equation}
  \label{eq:GaussianDef}
  \mathcal{P}(y_1,\ldots,y_{\rm p})\,{\rm d}y_1\cdots{\rm d}y_{\rm p}=
  \frac{1}{\sqrt{\left(2\pi\right)^p\det\left(\tens{\mathcal{M}}\right)}}\,
  {\rm e}^{-\mathcal{Q}}\,{\rm d}y_1\cdots{\rm d}y_{\rm p}
\end{equation}
with the quadratic form
\begin{equation}
  \label{eq:quadForm}
  \mathcal{Q}:=\frac{1}{2}\sum_{\rm i,j=1}^{p}\Delta y_{\rm i}
  \left(\tens{\mathcal{M}}^{-1}\right)_{\rm ij}
  \Delta y_{\rm j}\;,
\end{equation}
where $\tens{\mathcal{M}}$ is their \emph{covariance matrix} with
elements $\mathcal{M}_{ij}:=\langle\Delta y_i\Delta y_j\rangle$.
All statistical properties of a homogeneous Gaussian random field with
zero mean are fully characterised by the two-point correlation
function
$\zeta(\vec{r}_1,\vec{r}_2)=\zeta(|\vec{r}_1-\vec{r}_2|):=\langle
F(\vec{r}_1)F(\vec{r}_2)\rangle$ or equivalently its Fourier
transform, the power spectrum $P(k)$. In our case, this is the sum of
the matter fluctuation convergence power spectrum, $P_{\rm LSS}(k)$,
and observational noise, $P_{\rm n}(k)$.

Since we are interested in gravitational-lensing quantities such as
the convergence $\kappa$, we here consider two-dimensional Gaussian
random fields only. We adopt the formalism of \citet{BA86.1}, where
$F=\kappa$, $\eta_i=\partial_iF$ and
$\zeta_{ij}=\partial_i\partial_jF$ denote the convergence field and
its first and second derivatives, respectively.

\subsection{Definition of detections: a new up-crossing criterion}
\label{sec:modUpcross}

We define as \textit{detection} any contiguous area of the field
$\kappa$ which exceeds a given threshold,
$\kappa_\mathrm{th}=\mbox{S/N}\cdot\sigma_{\tilde{A}}$, determined by
the required signal-to-noise ratio, S/N, and the variance
$\sigma_{\tilde{A}}$ of an estimator $\tilde A$ (see
Eq.~\ref{eq:sigma_A}). This definition is widely used in surveys for
galaxy clusters or peak counts in weak-lensing surveys and can easily
be applied both to real data and Gaussian random fields.

\begin{figure}[t]
  \includegraphics[width=\hsize]{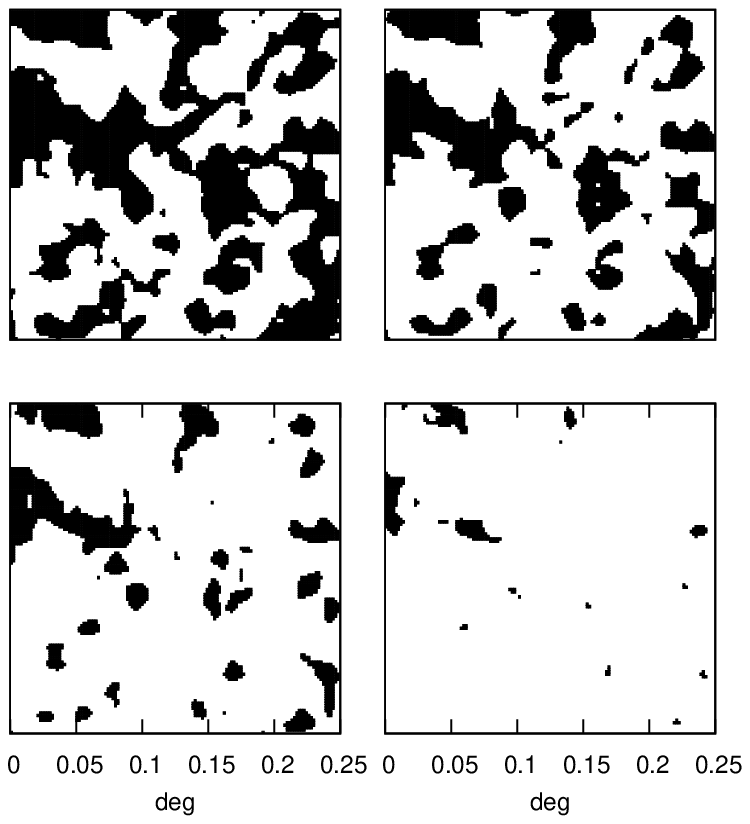}
  \includegraphics[width=\hsize]{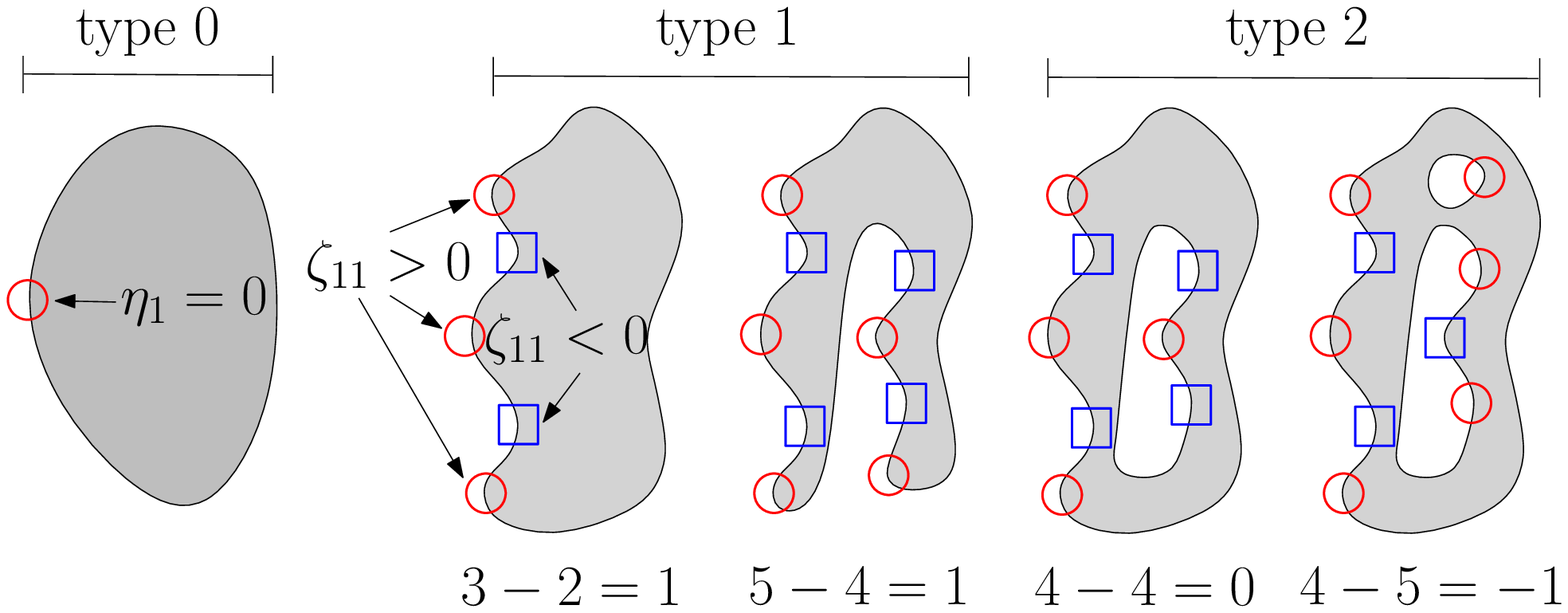}
  \caption{Weak lensing detection maps. The top four panels show the
    segmentation of a realistic weak-lensing S/N map for increasing
    thresholds: 0.1, 0.5, 1, and 2, respectively. The bottom panel
    sketches the three discussed detections types together with the
    points identified by the standard and the modified up-crossing
    criteria. Red circles and blue squares correspond to up-crossing
    points for which the second field derivatives are $\zeta_{11}>0$
    and $\zeta_{11}<0$, respectively.}
\label{fig:contours}
\end{figure}

Each detection is delimited by its contour at the threshold level
$\kappa_\mathrm{th}$. If this contour is approximately circular, 
it has a single point $\vec{r}_\mathrm{up}$, called up-crossing point,
  where the field is rising along the x-axis direction only, i.e.
where the field gradient has one vanishing and one positive component
(see the sketch for type-0 detections in the lower panel of
Fig.~\ref{fig:contours}),
\begin{equation}
  \label{eq:upcrossing}
  F(\vec{r}_\mathrm{up})=\kappa_\mathrm{th}\;,
  \quad
  \eta_1(\vec{r}_\mathrm{up})=0\;,
  \quad
  \eta_2(\vec{r}_\mathrm{up})>0\;.
\end{equation}
Since we assume $\kappa$ to be a homogeneous and isotropic random
field, the orientation of the coordinate frame is arbitrary and
irrelevant. The conditions expressed by Eq.~(\ref{eq:upcrossing})
define the so-called \textit{up-crossing criterion} which allows to
identify the detections and to derive their statistical properties,
such as their number counts, by associating their definition to the
Gaussian random field variables $F$, $\eta_1$ and $\eta_2$.

However, this criterion is prone to fail for low thresholds, where
detections tend to merge and the isocontours tend to deviate from the
assumed approximately circular shape. This causes detection numbers to
be overestimated at low cut-offs because each ``peninsula'' and
``bay'' of their profile (see type-1 in Fig.~\ref{fig:contours}) would
be counted as one detection.
We solve this problem by dividing the up-crossing points into those
with positive (red circles) and those with negative (blue squares)
curvature, $\zeta_{11}>0$ and $\zeta_{11}<0$ respectively. In fact,
for each detection, their difference is one (type-1) providing the
correct number count. The only exception is for those detections
containing one or more ``lagoons'' (type-2) since each of them
decreases the detection count by one. But since this is not a frequent
case and occurs only at very low cut-off levels, we do not consider
this case here.

\subsection{The number density of detections}

Once the relation between the detections and the Gaussian random
variables $\vec{y}=(\kappa_{\rm th},\eta_1,\eta_2,\zeta_{11})$ and
their constraints from Eq.~(\ref{eq:upcrossing}) together with
$\zeta_{11}>0$ or $\zeta_{11}<0$ are defined, we can describe their
statistical properties through the multivariate Gaussian probability
distribution given by Eq.~(\ref{eq:GaussianDef}) with the covariance
matrix
\begin{equation}
  \label{eq:corrMatrix}
  \tens{\mathcal{M}}=\left(\begin{array}{cccc}
      \sigma_0^2 & 0 & 0 & -\sigma_1^2/2 \\
      0 & \sigma_1^2/2 & 0 & 0 \\
      0 & 0 & \sigma_1^2/2 & 0 \\
      -\sigma_1^2/2 & 0 & 0 & 3\sigma_2^2/8 \\
    \end{array} \right)\;.
\end{equation}
This matrix differs from that derived by \citet{BA86.1} and
\citet{AN09.1} because we are dealing with a 2-dimensional rather than
a 3-dimensional field. Here, the $\sigma_j$ are the moments of the
power spectrum $P(k)$
\begin{equation}
  \label{eq:specMom}
  \sigma_j^2=\int\frac{k^{2j+1}\,\d k}{2\pi}P(k)\hat{W}^2(k)|\hat{Q}(k)|^2\;,
\end{equation}
where $\hat{W}(k)$ is the frequency response of the survey given by
its geometry (see Sec.~\ref{sec:windows}) and $\hat Q(k)$ is the
Fourier transform of the filter adopted for the weak lensing estimator
(see Sec.~\ref{sec:filters}). The determinant of $\tens{\mathcal{M}}$
is $(3\sigma_0^2\sigma_1^4\sigma_2^2-2\sigma_1^8)/32$ and
Eq.~(\ref{eq:quadForm}) can explicitly be written as
\begin{equation}
  \label{eq:quadFormKappa}
  \mathcal{Q}=\frac{1}{2}
  \left(
    \frac{2\vec{\eta}^2}{\sigma_1^2}+
    \frac{8\zeta_{11}^2\sigma_0^2+8\zeta_{11}\kappa\sigma_1^2+3\kappa^2\sigma_2^2}
    {3\sigma_0^2\sigma_2^2-2\sigma_1^4}
  \right)\;.
\end{equation}
Both $\kappa$ and $\eta_1$ can be expanded into Taylor series around
the points $\vec{r}_\mathrm{up}$ where the up-crossing conditions are
fulfilled,
\begin{equation}
  \label{eq:Taylor}
  \kappa(\vec{r})\approx
  \kappa_\mathrm{th}+
  \sum_{i=1}^2 \eta_i(\vec{r}-\vec{r}_\mathrm{up})_i\;,
  \quad
  \eta_1(\vec{r})\approx
  \sum_{i=1}^2\zeta_{1i}(\vec{r}-\vec{r}_\mathrm{up})_i\;,
\end{equation}
so that the infinitesimal volume element $\d\kappa\d\eta_1$ can be
written as $\d\kappa \d\eta_1=|\det \tens{J}| \d^2r$, where $\tens{J}$
is the \emph{Jacobian matrix},
\begin{equation}
  \label{eq:Jacobian}
  \tens{J}=\left(\begin{array}{cc}
      \partial\kappa/\partial x_1 & \partial\kappa/\partial x_2 \\
      \partial\eta_1/\partial x_1 & \partial\eta_1/\partial x_2 \\
    \end{array}\right)=
  \left(\begin{array}{cc}
      \eta_1 & \eta_2 \\
      \zeta_{11} & \zeta_{12} \\
    \end{array}\right)
\end{equation}
and $|\det\tens{J}|=|\eta_2\zeta_{11}|$ since $\eta_1=0$. The number
density of up-crossing points at the threshold $\kappa_\mathrm{th}$
with $\zeta_{11}>0$, and $\zeta_{11}<0$, $n^+$ and $n^-$ respectively,
can thus be evaluated as
\begin{equation}
  \label{eq:numPos}
  n^{\pm}(\kappa_\mathrm{th})=
  \pm
  \int\limits_0^\infty \d\eta_2
  \int\limits_0^\infty \d\zeta_{11}
  |\eta_2\zeta_{11}|\,
  \mathcal{P}
  \left(
    \kappa=\kappa_\mathrm{th},\eta_1=0,\eta_2,\zeta_{11}
  \right)\,,
\end{equation}
where $\mathcal{P}(\kappa,\eta_1,\eta_2,\zeta_{11})$ is the
multivariate Gaussian defined by Eq.~(\ref{eq:GaussianDef}) with
$p=4$, the correlation matrix (\ref{eq:corrMatrix}), and the quadratic
form (\ref{eq:quadFormKappa}). Both expressions can be integrated
analytically and their difference,
$n_\mathrm{det}(\kappa_\mathrm{th})=n^{+}(\kappa_\mathrm{th})-n^{-}(\kappa_\mathrm{th})$,
as explained in Sect.~\ref{sec:modUpcross}, returns the number density
of detections $n_\mathrm{det}$ above the threshold
$\kappa_\mathrm{th}$,
\begin{equation}
 \label{eq:detections}
 n_\mathrm{det}(\kappa_\mathrm{th})=
 \frac{1}{4\sqrt{2}\pi^{3/2}}
 \left(
   \frac{\sigma_1}{\sigma_0}
 \right)^2
 \frac{\kappa_\mathrm{th}}{\sigma_0}
 \exp
 \left(
   -\frac{\kappa_\mathrm{th}^2}{2\sigma_0^2}
 \right)\;.
\end{equation}
Note how the dependence on $\sigma_2$ drops out of the difference
$n^+-n^-$, leading to a very simple result.

For completeness we report the number density estimate also for the
classical up-crossing criterion, Eq.~(\ref{eq:upcrossing}) only, where
the constraint on the second derivative of the field, $\zeta_{11}$, is
not used,
\begin{eqnarray}
  \label{eq:upcrossingAnal}
  n_\mathrm{up}(\kappa_\mathrm{th})
  &=&
  \frac{1}{4\sqrt{2}\pi^2\sigma_0}
  \left(
    \frac{\sigma_1}{\sigma_0}
  \right)^2
  \exp
  \left(
    -\frac{\kappa_\mathrm{th}^2}{2\sigma_0^2}
  \right) \\
  &\times&
  \left[
    \exp
    \left(
      -\frac{\kappa_\mathrm{th}^2\sigma_1^4}{\sigma_0^2\gamma^2}
    \right)
    \sigma_0\gamma+\sqrt{\pi}\kappa_\mathrm{th}\,\mathrm{erf}
    \left(
      \frac{\kappa_\mathrm{th}\sigma_1^2}{\sigma_0\gamma}
    \right)
  \right]\;, \nonumber
\end{eqnarray}
with $\gamma:=\sqrt{3\sigma_0^2\sigma_2^2-2\sigma_1^4}$.
This number density converges to the correct value $n_\mathrm{det}$
for $\kappa_\mathrm{th}\rightarrow\infty$, i.e. large thresholds,
because $\mathrm{erf}(x)\rightarrow1$ and
$\exp(-x^2)\rightarrow0$. This reflects the fact that, for large
thresholds, the detection shapes become fully convex and any issues
with more complex shapes disappear.

\section{Analytic predictions vs. numerical simulations}

We now compare the number counts of detections predicted by our
analytic approach with those resulting form the analysis of synthetic
galaxy catalogues produced with numerical ray-tracing simulations.

\begin{figure}[!t]
  \includegraphics[angle=-90,width=\hsize]{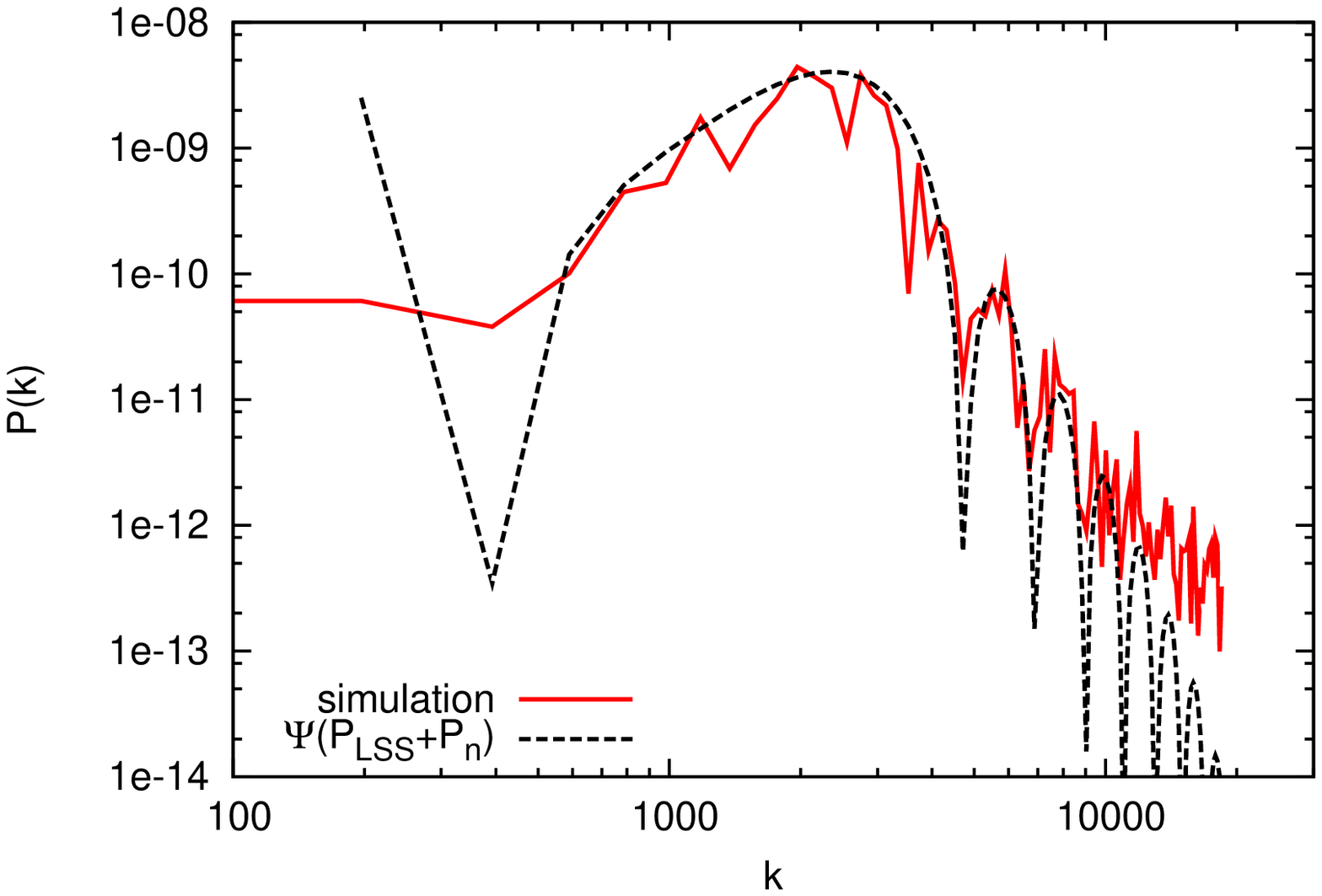}
  \includegraphics[angle=-90,width=\hsize]{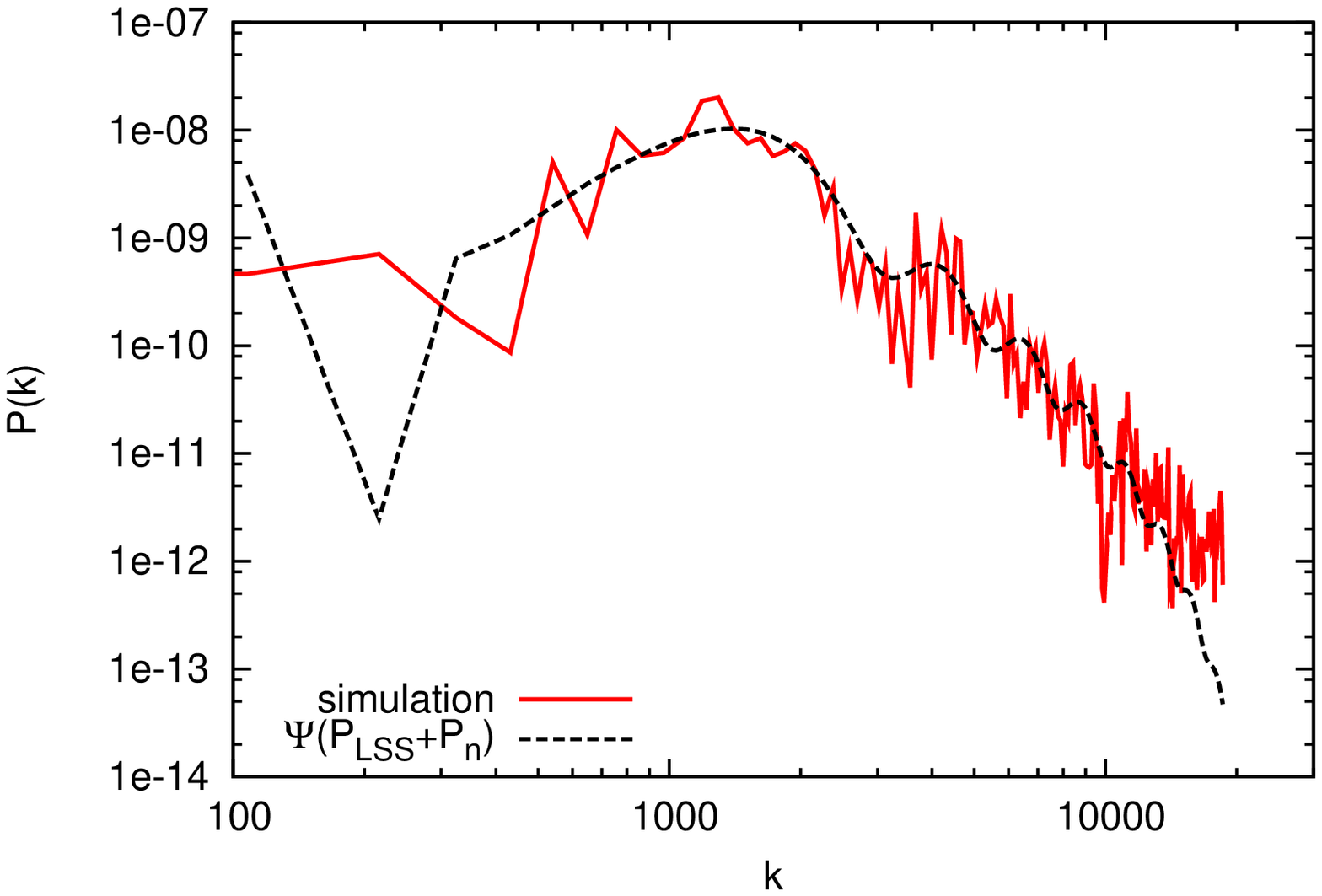}
  \includegraphics[angle=-90,width=\hsize]{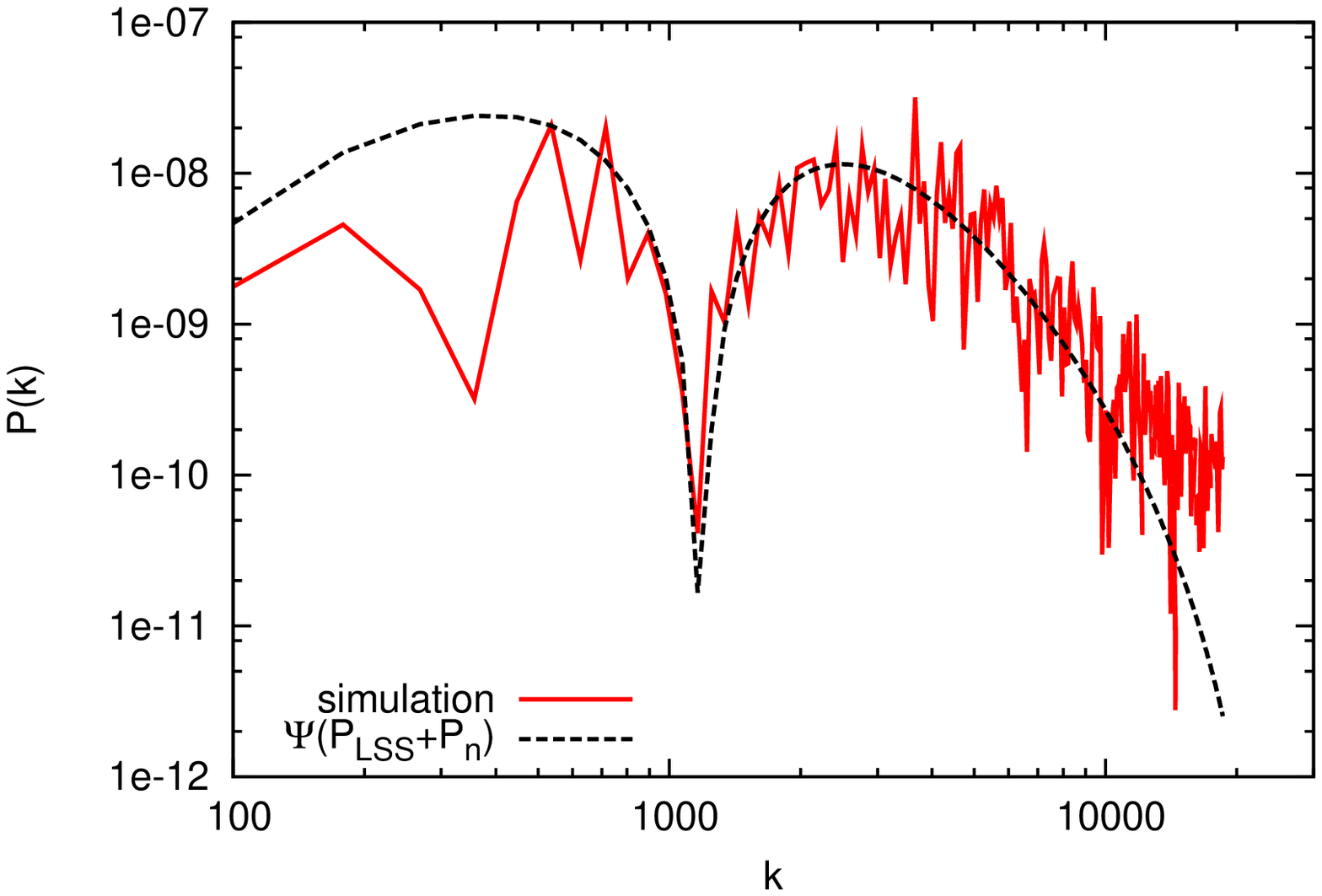}
  \caption{Comparison between the power spectrum measured from the
    analysis of shear catalogues derived from numerical simulations
    (heavy line) and the prediction resulting from the \cite{PE96.1}
    power spectrum convolved with the survey frequency response and
    the adopted filter. We show from top to bottom the results for the
    polynomial filter with $r_s=5\farcm5$, the hyperbolic-tangent
    filter with $r_s=10'$ and the optimal filter with the cluster
    model scale radius set to $r_s=1'$.}
\label{fig:spectra}
\end{figure}

\subsection {Numerical simulations}\label{sec:numericals}

We use a hydrodynamical, numerical $N$-body simulation carried out
with the code {\small GADGET-2} \citep{SP05.1}. We briefly summarise
its main characteristics here and refer to \citet{BO04.1} for a more
detailed discussion. The simulation represents a concordance
$\Lambda$CDM model, with dark-energy, dark-matter and baryon density
parameters $\Omega_{\Lambda}=0.7$, $\Omega_{\rm m}=0.3$ and
$\Omega_{\rm b}=0.04$, respectively. The Hubble constant is
$H_0=100\,h\,\mathrm{km\,s^{-1}\,Mpc^{-1}}$ with $h=0.7$, and the
linear power spectrum of the matter-density fluctuations is normalised
to $\sigma_8=0.8$. The simulated box is a cube with a side length of
$192\,h^{-1}$Mpc, containing $480^3$ dark-matter particles with a mass
of $6.6\times10^9\,h^{-1}M_{\odot}$ each and an equal number of gas
particles with $8.9\times10^8\,h^{-1}M_{\odot}$ each. Thus, halos of
mass $10^{13}\,h^{-1}M_\odot$ are resolved into several thousands of
particles. The physics of the gas component includes radiative
cooling, star formation and supernova feedback, assuming zero
metallicity.

This simulation is used to construct backward light cones by stacking
the output snapshots from $z=1$ to $z=0$. Since the snapshots contain
the same cosmic structures at different evolutionary stages, they are
randomly shifted and rotated to avoid repetitions of the same cosmic
structures along one line-of-sight. The light cone is then sliced into
thick planes, whose particles are subsequently projected with a
triangular-shaped-cloud scheme \citep[TSC,][]{HO88.1} on lens planes
perpendicular to the line-of-sight. We trace a bundle of
$2048\times2048$ light rays through the light cone which start
propagating at the observer into directions on a regular grid of $4.9$
degrees on each side. The effective resolution of this ray-tracing
simulation is of $1'$ \citep[for further detail, see][]{PA07.1}.

The effective convergence and shear maps obtained from the ray-tracing
simulations are used to lens a background source population according
to Eq.~(\ref{eq:epstr}). Galaxies are randomly distributed on the lens
plane at $z=1$ with a number density of $n_{\rm g}=30$ arcmin$^{-2}$
and have intrinsic random ellipticities drawn from the distribution
\begin{equation}
  p(\epsilon_{\rm s})=
  \frac{\exp{|(1-|\epsilon_{\rm s}|^2)/\sigma_{\epsilon}^2|}}
  {\pi\sigma_{\epsilon}^2|\exp{(1/\sigma_{\epsilon}^2)}-1|}\;,
\end{equation}
where $\sigma_{\epsilon}=0.25$.

Synthetic galaxy catalogues produced in this way are finally analysed
with the aperture mass (Eq.~\ref{eq:A}) evaluated on a regular grid of
$512\times 512$ positions covering the entire field-of-view of the
light cone. All three filters presented in \S~\ref{sec:filters} were
used with three different scales: the polynomial filter with $r_{\rm
  s}=2\farcm75,\ 5\farcm5,$ and $11'$, the hyperbolic-tangent filter with
$r_{\rm s}=5',\ 10',$ and $20'$, and the optimal filter with scale radii
of the cluster model set to $r_s=1',\ 2',$ and $4'$. These scales
  are chosen such that the effective scales of the different filters
  can be compared. For a statistical analysis of the weak-lensing
detections and their relation to the numerical simulations structures,
see \cite{PA07.1}.

\begin{figure}[!t]
  \includegraphics[angle=-90,width=\hsize]{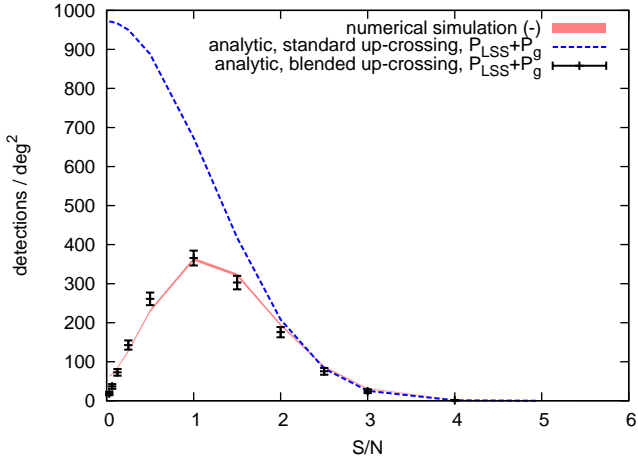}
  \caption{Number of peaks detected in the numerical simulation
    (shaded area) compared to the prediction obtained with the
    proposed method both with the original up-crossing criterion
    (dashed line) and with the new blended up-crossing criterion
    (points with error bars). The standard up-crossing criterion is a
    good approximation for high signal-to-noise ratios but fails for
    lower S/N, which are well described by the new version.}
\label{fig:upcrossing}
\end{figure}

\subsection{Accounting for the geometry of surveys: the window function}
\label{sec:windows}

Our analytic predictions for the number density of detections accounts
for the full survey geometry, through the frequency response
$\hat{W}(k)$ in Eq.~(\ref{eq:specMom}). Any survey geometry can be
considered but, for sake of simplicity, we evaluate our estimates for
a square-shaped field-of-view without gaps. Under these assumptions,
the frequency response, $W(k)$, is the product of a high-pass filter
suppressing the scales larger than the light cone's side length
$L_{\rm f}=2\pi/k_\mathrm{f}=4.9\;\mbox{deg}$,
\begin{equation}
  \hat{W}^2_\mathrm{f}(k)=
  \exp\left(-\frac{k_\mathrm{f}^2}{k^2}\right)\;,
\end{equation}
a low-pass filter imposed by the average separation $d=2\pi/k_{\rm
  g}=n_{\rm g}^{-1/2}=0\farcm18$ between the galaxies,
\begin{equation}
  \hat{W}_\mathrm{g}^2(k)=
  \exp\left(-\frac{k^2}{k_\mathrm{g}^2}\right)\;,
\end{equation}
and a low-pass filter related to the resolution
$R_\mathrm{pix}=0\farcm57$ used to sample the sky with the weak
lensing estimator $\tilde A$ of Eq.~(\ref{eq:A}),
\begin{equation} \label{eq:pixelsSph}
  \hat{W}_\mathrm{pix}^2(k)=
  \frac{2\,\sqrt{\pi}}{kR_\mathrm{pix}}\,
  \mathrm{J}_1\left(\frac{kR_\mathrm{pix}}{\sqrt{\pi}}\right)\;,
\end{equation}
where $\mathrm{J}_1(x)$ is the cylindrical Bessel function of order
one. The latter function is a circular step function
covering the same area as a square-shaped pixel of size
$R_\mathrm{pix}$. The square shapes of the field-of-view and the
pixels could be better represented by the product of two step
functions in both the $x$- and $y$-direction, but the low gain in
accuracy does not justify the higher computational cost.
Finally, for the comparison with our numerical ray-tracing simulation,
we have to account for its resolution properties which act on the
convergence power spectrum only by convolving $P_\kappa$ with a
low-pass filter
\begin{equation}
  \label{eq:finiteMass}
  \hat{W}_\mathrm{s}^2(k)=
  \exp\left(-\frac{k^2}{k_\mathrm{s}^2}\right)\;,
\end{equation}
where $k_\mathrm{s}=2\pi/1\;\mbox{arcmin}^{-1}$ as discussed in
\S~\ref{sec:numericals}.

Note that when relating the detection threshold to the signal-to-noise
ratio S/N according to the estimator variance using
Eq.~(\ref{eq:sigma_A}) and
$\kappa_\mathrm{th}=\mbox{S/N}\cdot\sigma_{\tilde{A}}$, all filters
mentioned are used except for $\hat W_{\rm pix}$, which, of course,
does not affect the estimator variance.

\subsection {Comparison with numerical simulations}
\label{sec:comparison}

\begin{figure*}[!t]
  \includegraphics[angle=-90,width=0.32\hsize]{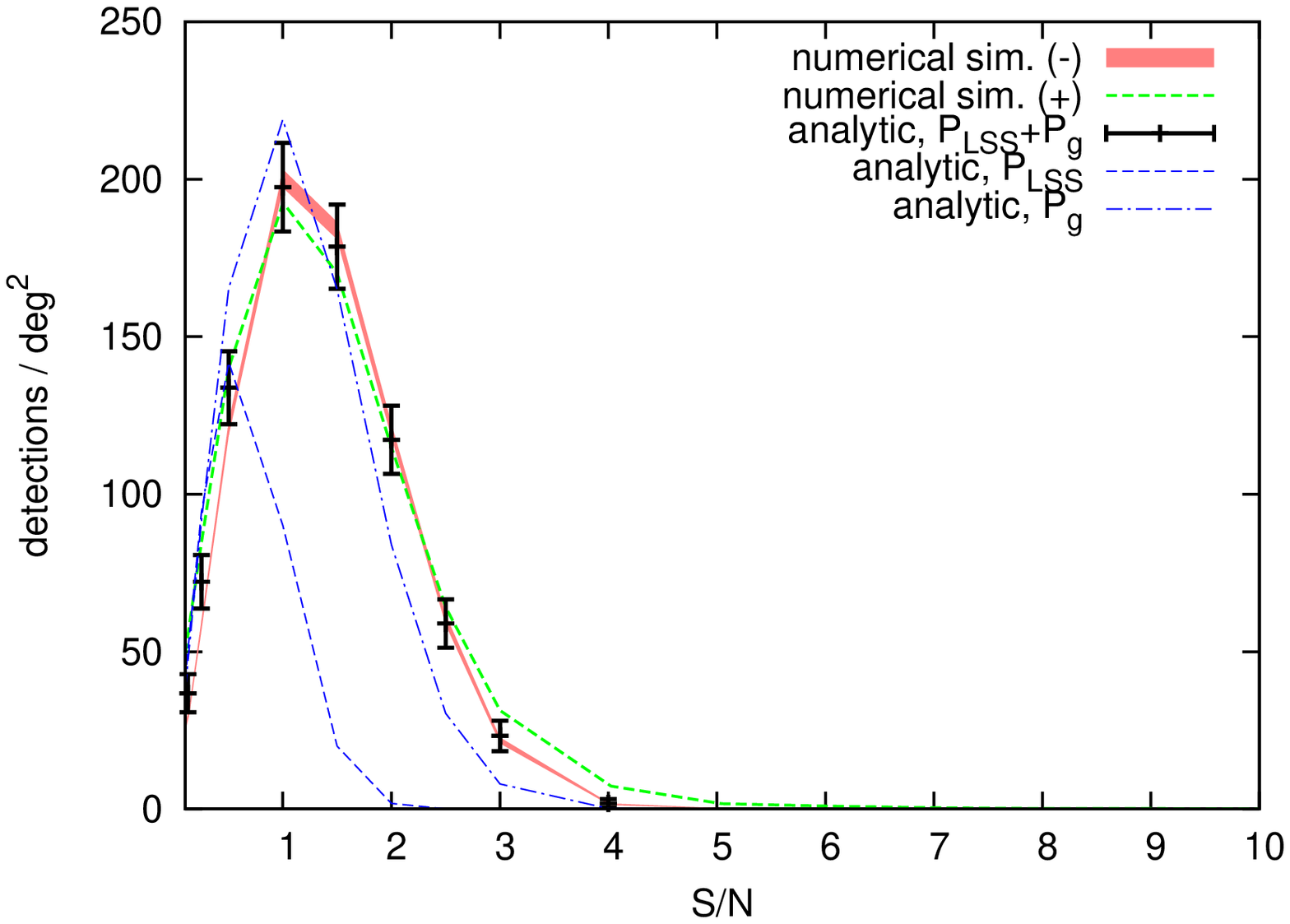}\hfill
  \includegraphics[angle=-90,width=0.32\hsize]{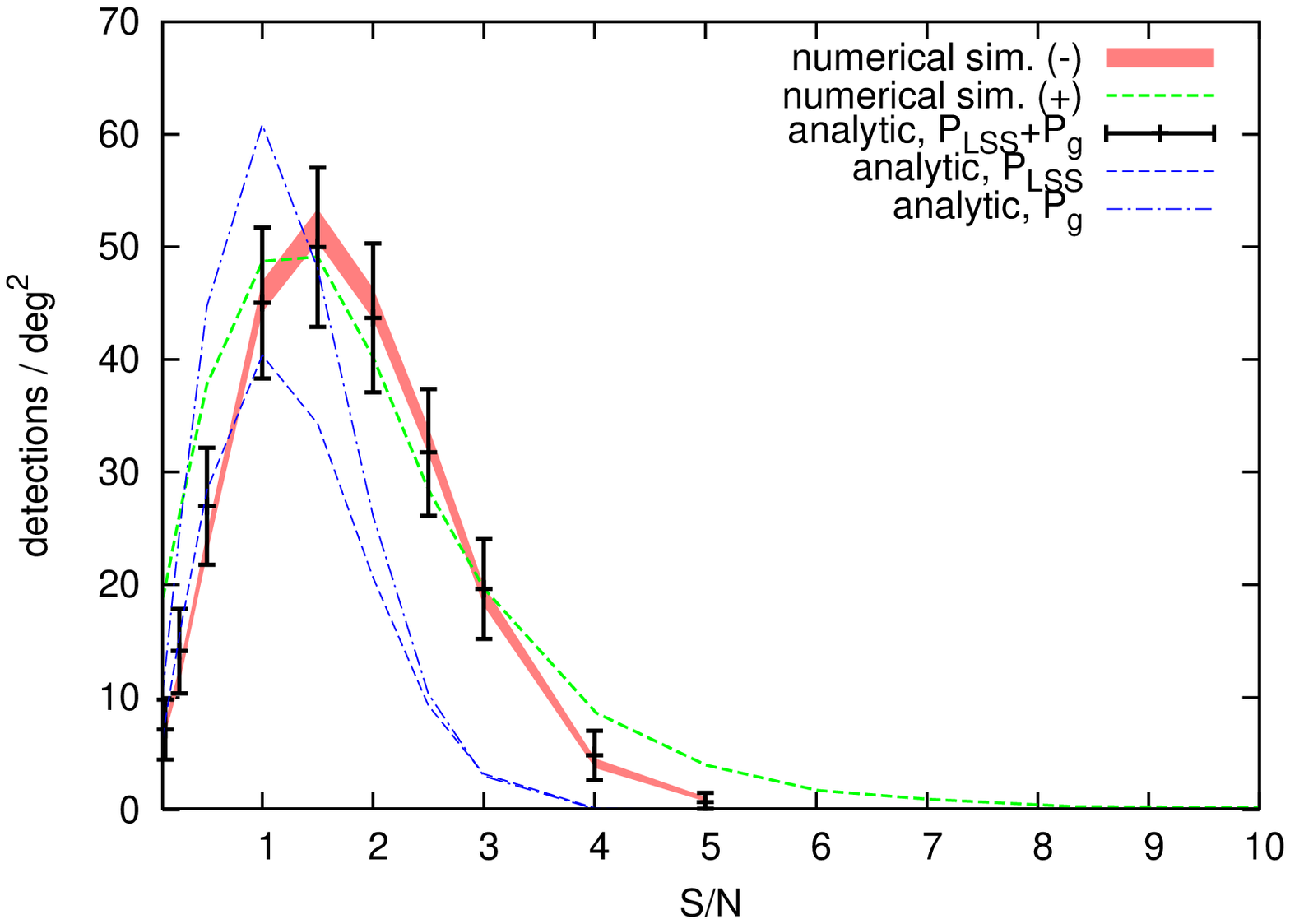}\hfill
  \includegraphics[angle=-90,width=0.32\hsize]{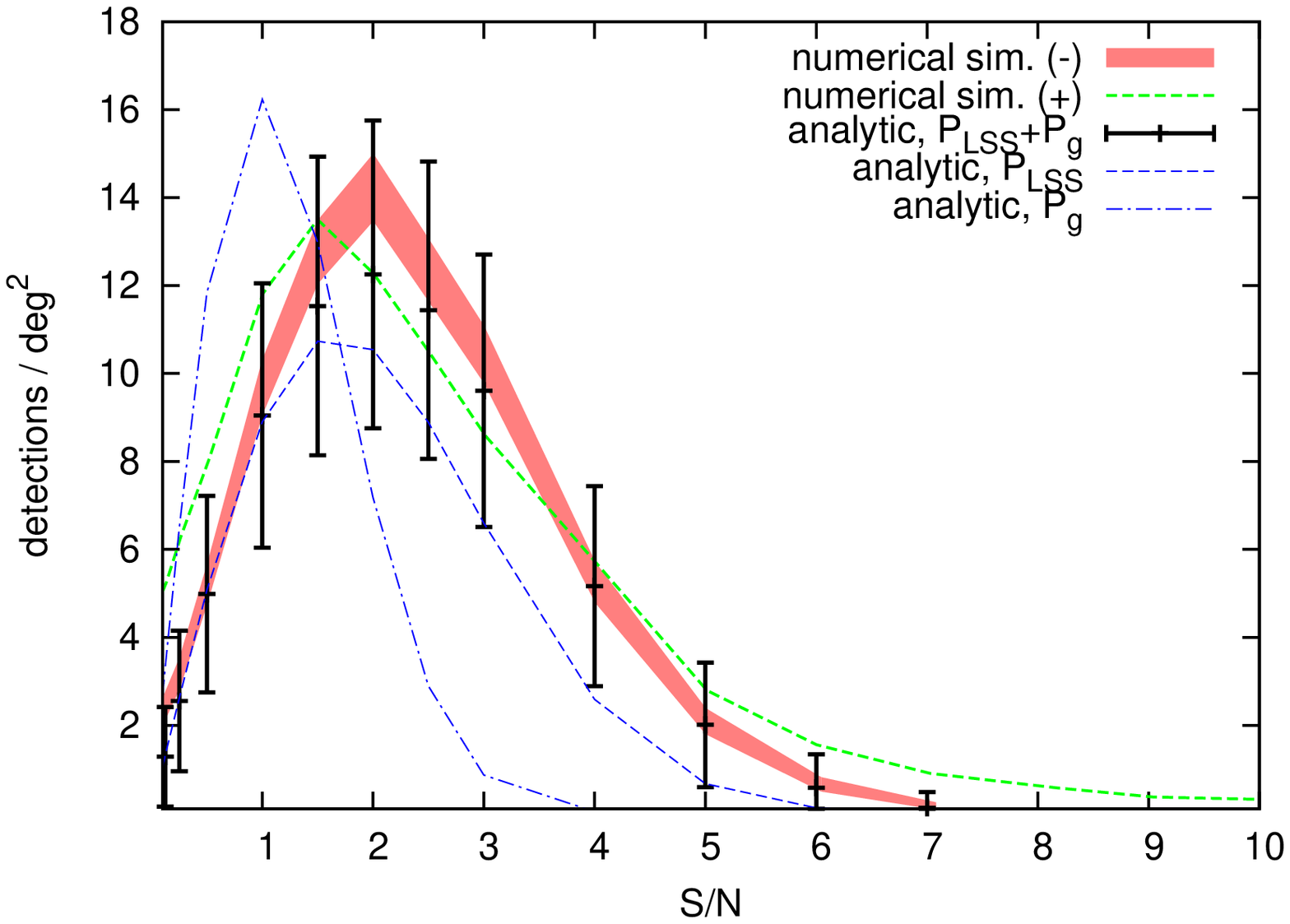}\\
  \includegraphics[angle=-90,width=0.32\hsize]{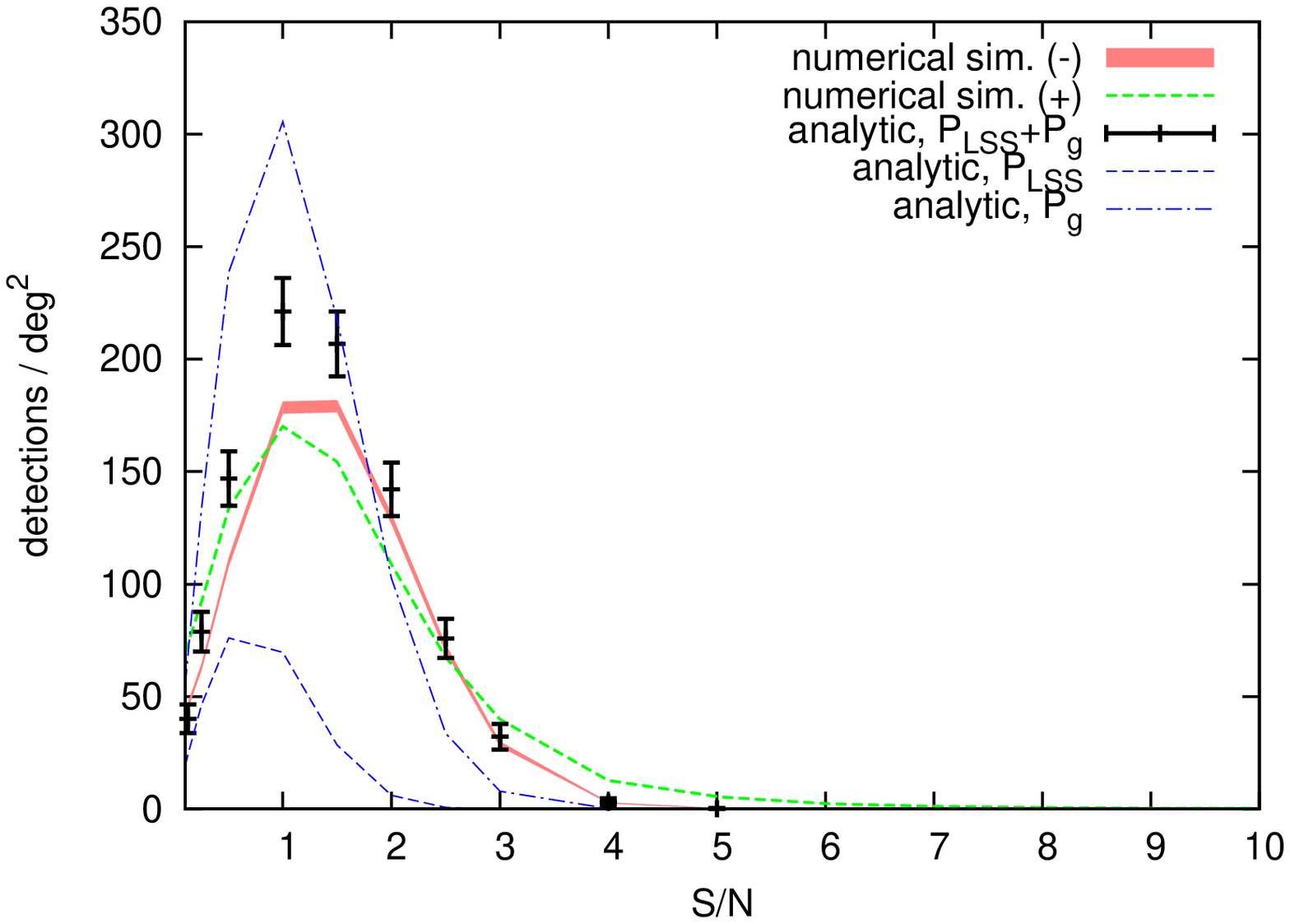}\hfill
  \includegraphics[angle=-90,width=0.32\hsize]{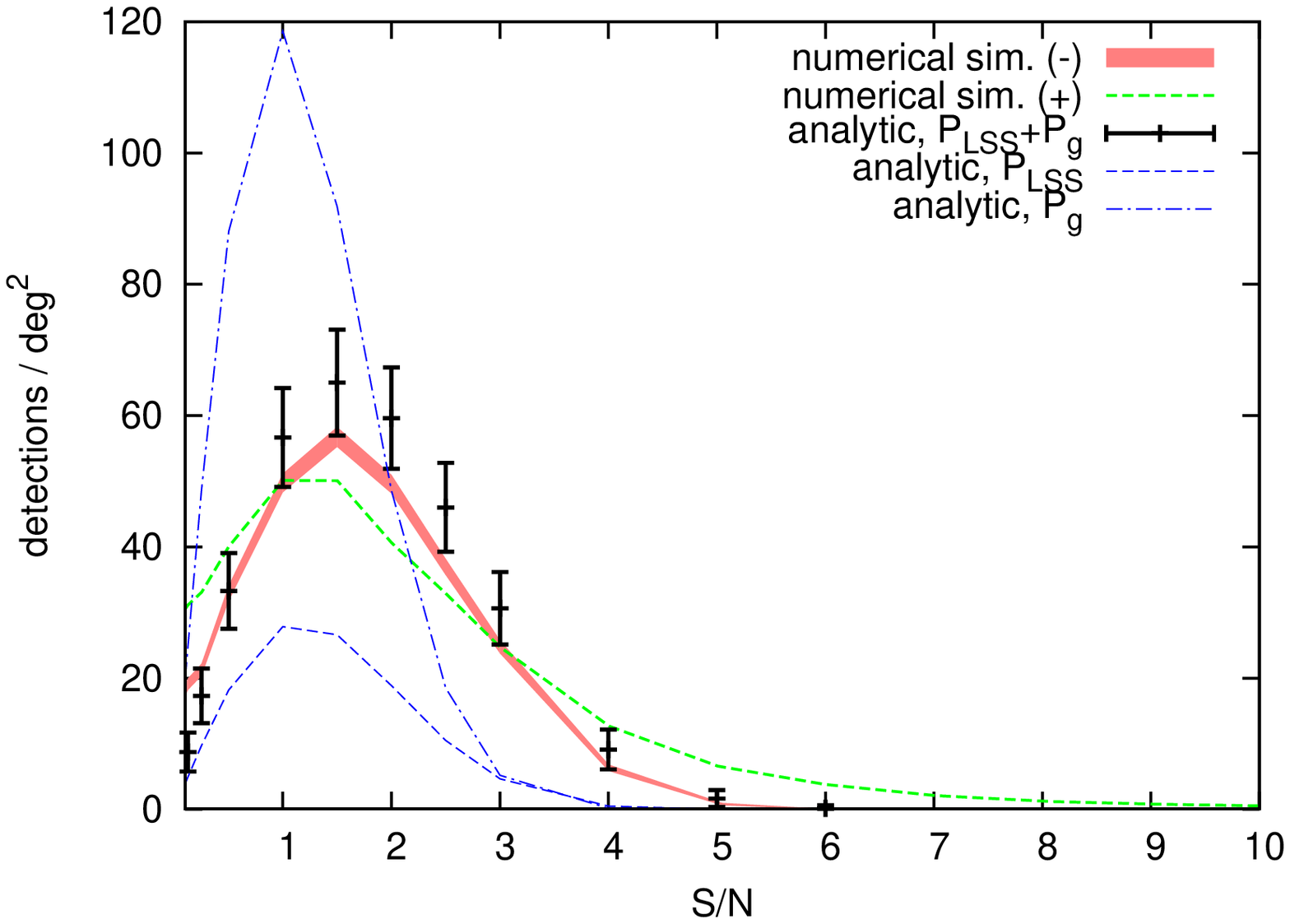}\hfill
  \includegraphics[angle=-90,width=0.32\hsize]{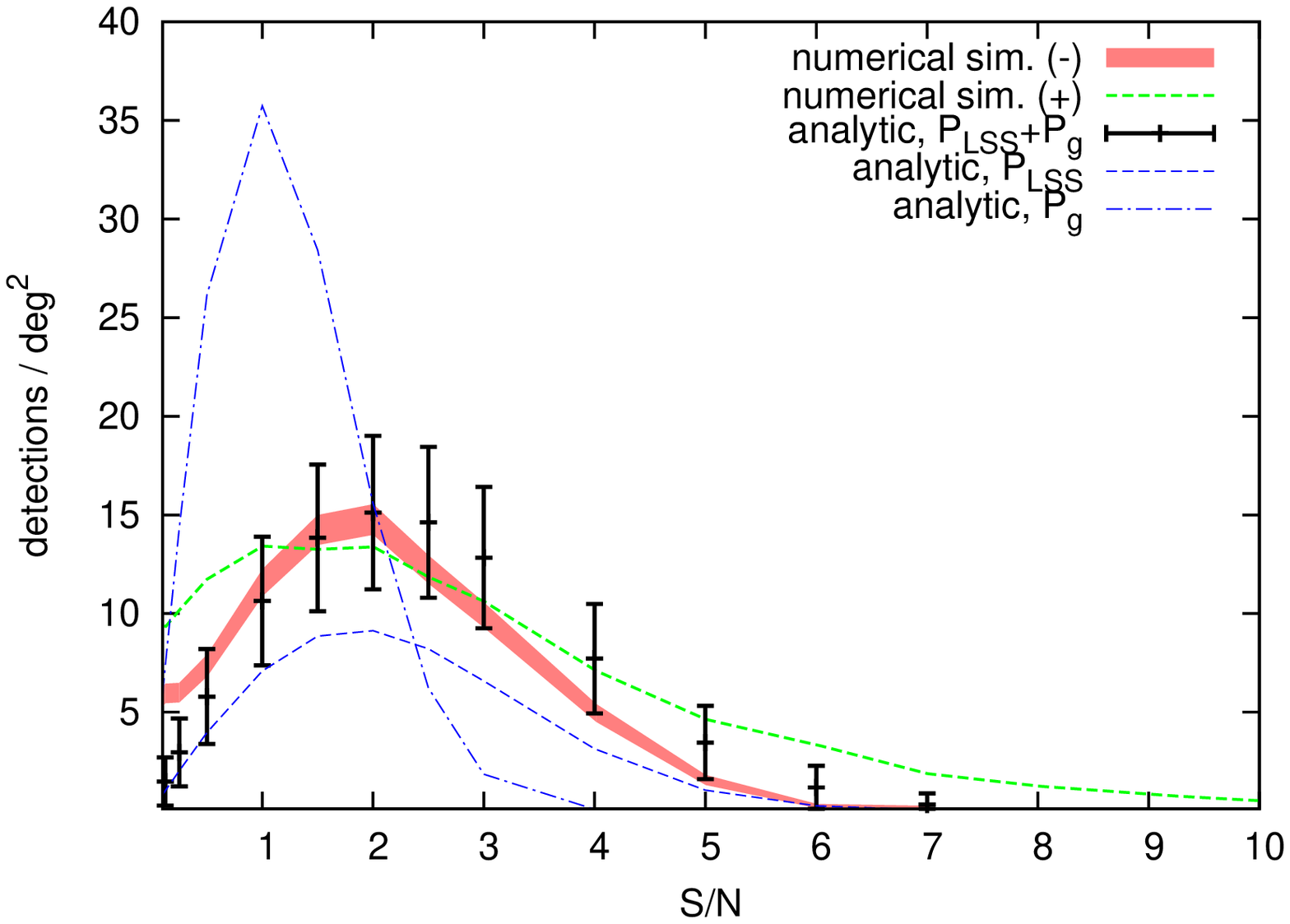}\\
  \includegraphics[angle=-90,width=0.32\hsize]{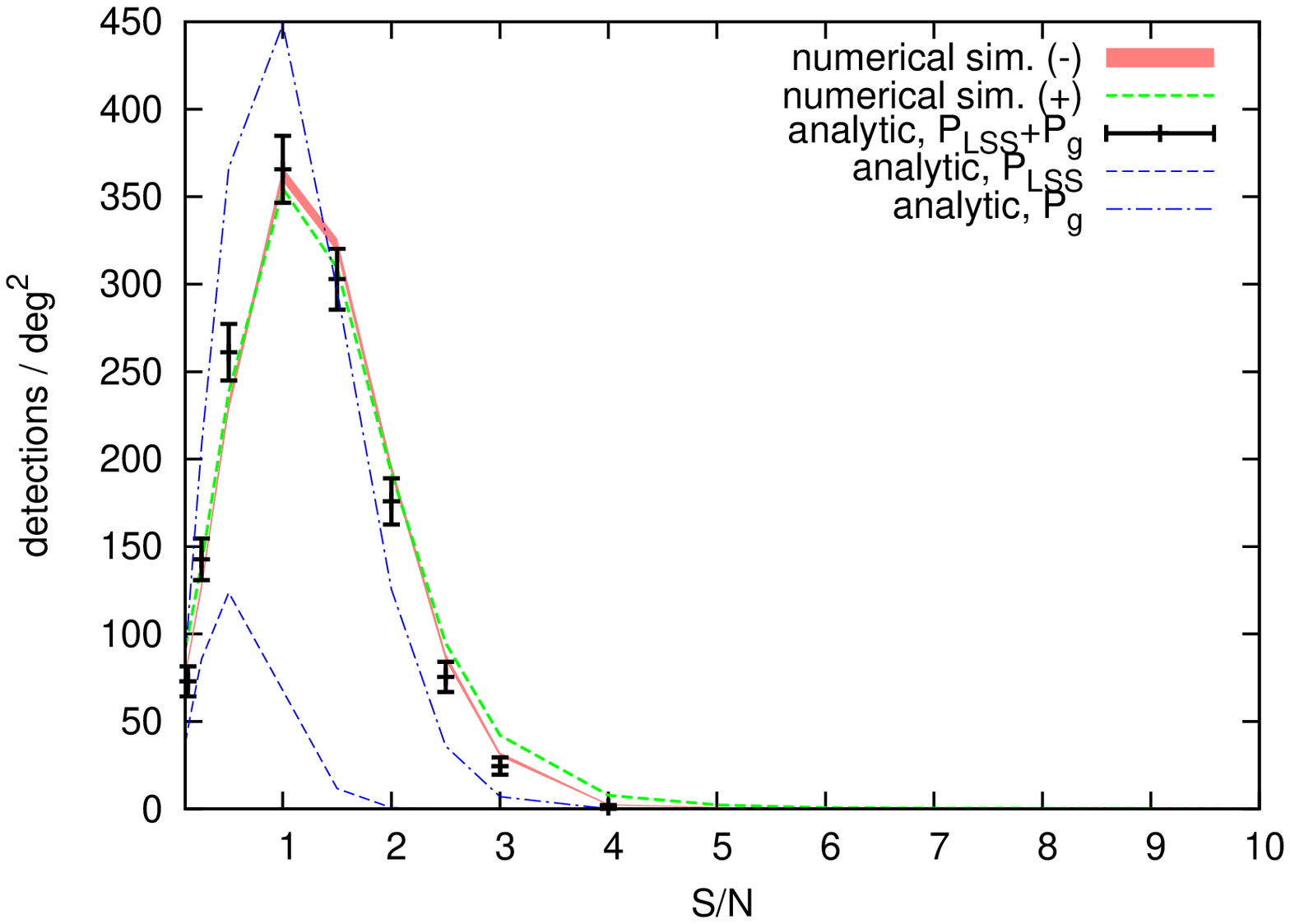}\hfill
  \includegraphics[angle=-90,width=0.32\hsize]{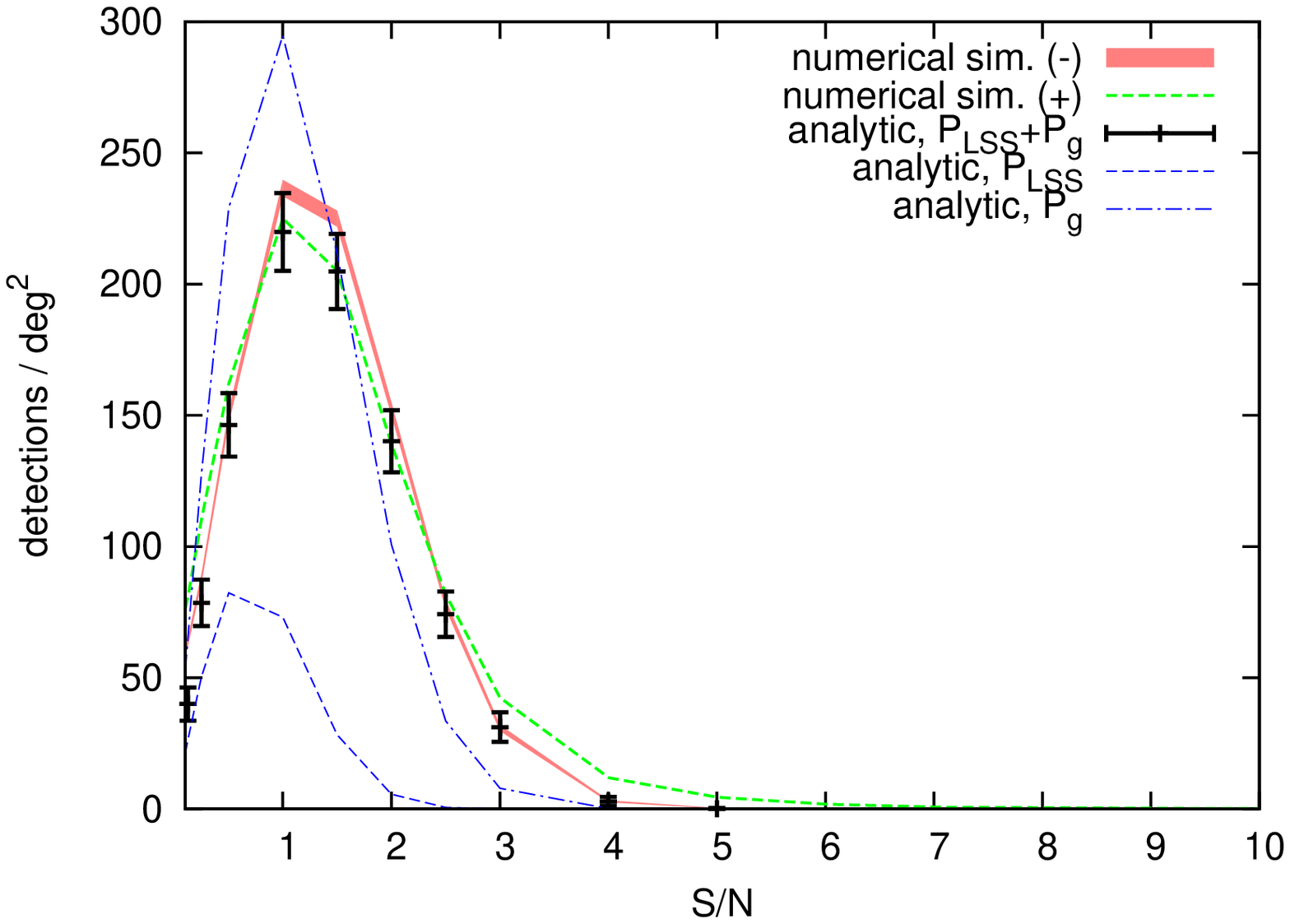}\hfill
  \includegraphics[angle=-90,width=0.32\hsize]{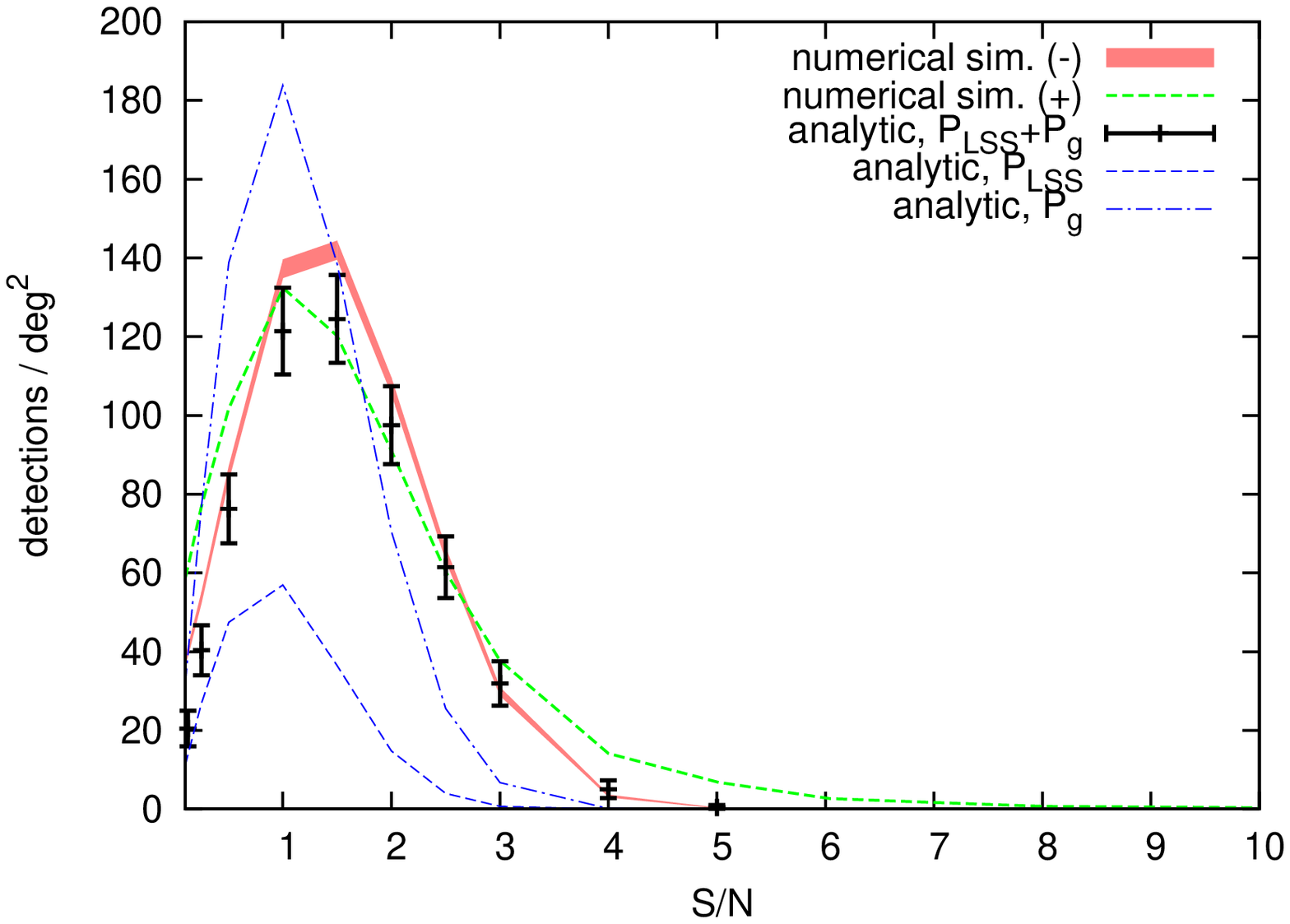}
  \caption{Number of weak lensing peaks, shown as a function of the
    signal-to-noise ratio, predicted with the analytic method
    presented here for the \cite{SC98.2}, the \cite{SC04.2} and
      the \cite{MAT04.2} filters from top to bottom, and increasing filter
    radii from left to right. The number counts generated by
    the intrinsic galaxy noise alone, $P_{\rm g}$, and the LSS alone,
    $P_{\rm LSS}$, are also shown. Numbers refer to a survey of one
    square degree with a galaxy number density of
    $n_{\rm g}=30\ \mbox{arcmin}^{-1}$ and an intrinsic shear dispersion of
    $\sigma_{\rm g}=0.25$. The results are compared with the number counts
    of positive as well as negative peaks detected based on the
    synthetic galaxy catalogues from the numerical simulation.}
\label{fig:numberSim}
\end{figure*}

We first compare in Fig.~\ref{fig:spectra} the expected filtered power
spectra for the convergence and the intrinsic noise of the galaxies,
$P |\hat{W}\hat{Q}|^2$, assumed in Eq.~(\ref{eq:specMom}), with those
measured from the analysed synthetic galaxy catalogue. All main
features are very well reproduced. Only at high frequencies the
assumed power spectra drop slightly more steeply than measured in the
numerical simulations.

A comparison of the original up-crossing criterion with the new
blended up-crossing criterion presented here is shown in
Fig.~\ref{fig:upcrossing} together with the number counts obtained
from the numerical simulations. Only the result for the optimal filter
with $r_{\rm s}=1'$ is shown for clarity. As expected, the two
criteria agree very well for high signal-to-noise ratios since the
detections are mostly of type-0, i.e.~approximately circular, as shown
in the left panel of Fig.~\ref{fig:contours}, while the merging of
detections at lower signal-to-noise ratios is correctly taken into
account only by our new criterion. Note that our analytic approach
approximates the data as random Gaussian fields, very well
representing noise and LSS but not non-linear objects such as galaxy
clusters. Thus, particular care has to be taken when comparing the
predicted number counts with real or simulated data by modelling the
non-linear structures, which is difficult and uncertain, or by
avoiding their contribution in first place. We follow the latter
approach by counting the \textit{negative} instead of the positive
peaks found in the galaxy catalogues. In fact, massive halos
contribute only positive detections in contrast to the LSS and other
sources of noise which equally produce positive and negative
detections with the same statistical properties.

{\it This is an interesting application for cosmology, where a robust
  prediction of number counts in the linear regime only can be
  directly compared to data or, turning the argument around, can be
  used to statistically correct the halo number counts by using the
  data only. In fact, the difference of the positive and negative
  detection counts is caused by non linear structures only and their
  Poisson fluctuations.}

Our analytic predictions for all filters and both positive and
negative detection counts resulting from the synthetic galaxies
catalogue from the numerical simulation are shown in
Fig.~\ref{fig:numberSim}. The high signal-to-noise ratio tail caused
by the nonlinear structures is present only in the positive detection
counts, as expected. The agreement with the negative detections is
within the 1-$\sigma$ error bars (considering the uncertainties for a
one square degree survey) except for \cite{SC04.2} filter and the
optimal filter, both with a scale of 5 arcmin, which are compatible
only at a 2-$\sigma$ level. It is plausible that these small deviations
are caused by the ignored correlation between the pixels which, in the
noise evaluation, are treated as independent. The filter radii are in
fact larger than the pixel separation, so that different pixels have
some common galaxies.

We finally compare the contribution of the LSS and the noise to the
total signal by treating them separately. Their number counts are
plotted with dashed lines in Fig.~\ref{fig:numberSim}. All filters
show an unsurprisingly large number of detections caused by the noise
up to signal-to-noise rations of 3 and a number of detections caused
by the LSS increasing with the filter scale except for the optimal
filter, which always suppresses their contribution to a negligible
level. Thus, the LSS contaminates halo catalogues selected by weak
lensing up to signal-to-noise ratios of $4-5$ if its contribution is
ignored in the filter definition.
Note that the total number of detections can be obtained only by
counting the peaks from the total signal, i.e.~LSS plus noise, and not
by adding the peaks found in the two components separately, because
the blending of peaks is different for the two cases.

\section{Conclusion}

We have developed an analytic method for predicting peak counts in
weak-lensing surveys, based on the theory of Gaussian random fields
\citep{BA86.1}. Peaks are typically detected in shear fields after
convolving them with filters of different shapes and widths. We have
taken these into account by first filtering the assumed Gaussian
random field appropriately and then searching for suitably defined
peaks. On the way, we have argued for a refinement of the up-crossing
criterion for peak detection which avoids biased counts of detections
with low signal-to-noise ratio, and implemented it in the analytic
peak-count prediction. Peaks in the non-linear tail of the shear
distribution are underrepresented in this approach because they are
highly non-Gaussian, but our method is well applicable to the
prediction of spurious counts, and therefore to the quantification of
the background in attempts to measure number densities of dark-matter
halos. We have compared our analytic prediction to peak counts in
numerically simulated, synthetic shear catalogues and found agreement
at the expected level.

Our main results can be summarised as follows:

\begin{itemize}
\item The shape and size of the filter applied to the shear field has
  a large influence on the contamination by spurious detections. For
  the optimal filter, the contribution by large-scale structures is
  low on all filter scales, while they typically contribute
  substantially for other filters.
\item Shape and shot noise due to the background galaxies used to
  measure the shear from are a large source of spurious peak counts
  for all filters, and the dominant source for the optimal filter.
\item Taken together, large-scale structure and galaxy noise
  contribute the majority of detections up to signal-to-noise ratios
  between 3--5. Only above this level, detections due to real
  dark-matter halos begin dominating.
\item The optimal filter allows the detection of $\sim$30--40 halos
  per square degree at signal-to-noise ratios high enough for
  suppressing all noise contributions. For the other filters, this
  number is lower by almost an order of magnitude.
\end{itemize}

Our conclusions are thus surprisingly drastic: peak counts in
weak-lensing surveys are almost exclusively caused by chance
projections in the large-scale structure and by galaxy shape and shot
noise unless only peaks with high signal-to-noise ratios are
counted. With typical filters, only a few detections per square degree
can be expected at that level, while the optimal filter returns up to
an order of magnitude more.

\acknowledgements{This work was supported by the Transregional
  Collaborative Research Centre TRR~33 (MM, MB) and grant number
  BA~1369/12-1 of the Deutsche Forschungsgemeinschaft, the Heidelberg
  Graduate School of Fundamental Physics and the IMPRS for Astronomy
  \& Cosmic Physics at the University of Heidelberg.}

\bibliographystyle{aa}

\appendix

\section{Forecast for different weak lensing surveys}

For convenience, we evaluate here the expected number density of peak
counts for a collection of present and future weak-lensing surveys. To
give typical values, we assumed for all of them a square-shaped field
of view, a uniform galaxy number density and no gaps for two main
reasons. First, their fields-of-view are typically very large and thus
do not affect the frequencies relevant for our evaluation. Second, the
masking of bright objects can be done in many different ways which
cannot be considered in this paper in any detail. Finally we fixed the
sampling scale, described by Eq.~(\ref{eq:pixelsSph}), to be 5 times
smaller than the typical filter scale in order to avoid undersampling,
i.e. such that the high frequency cut-off is imposed by the filters
themselves. The results are shown in Tab.~\ref{tab:surveys} together
with the number counts obtained with a simple Gaussian filter, usually
used together with the Kaiser \& Squires shear inversion algorithm
\citep{KA93.4}.

\begin{table*}
  \label{tab:surveys}
  \centering
  \caption{Expected number counts of peak detections per square degree for $\mathrm{S/N}=1,3,5$ and for several present and future weak-lensing surveys with different intrinsic ellipticity dispersion, $\sigma_\epsilon$, and galaxy number density, $n_{\rm g}$, per arcmin$^{2}$. For each filter, we used three different scales, namely \emph{$Q_{poly}$:} scale-1=$2\farcm75$, scale-2=$5\farcm5$, scale-3=$11'$; \emph{$Q_{tanh}$:} scale-1=$5'$, scale-2=$10'$, scale-3=$20'$; \emph{$Q_{opt}$:} scale-1=$10^{14}\,\mathrm{M}_\odot$ and scale-2=$5\times10^{14}\,\mathrm{M}_\odot$. \emph{$Q_{gaus}$ (Gaussian FWHM):} scale-1=$1'$, scale-2=$2'$, scale-3=$5'$.}
  \begin{tabular}{|c|c c c|c c c|c c c|c c c|}
    \hline
    \textbf{Pan-STARRS} & \multicolumn{3}{c|}{\textbf{$Q_{poly}$}} & \multicolumn{3}{c|}{\textbf{$Q_{tanh}$}} & \multicolumn{3}{c|}{\textbf{$Q_{opt}$}} & \multicolumn{3}{c|}{\textbf{$Q_{gaus}$}} \\ 
    $\sigma_\epsilon=0.3$, $n_{\rm g}=5$ & 1 & 3 & 5 & 1 & 3 & 5 & 1 & 3 & 5 & 1 & 3 & 5 \\ \hline
    scale-1 & 207.7 & 8.127 & 0.002 & 252.7 & 8.824 & 0.002 & 186.3 & 6.243 & 0.001 & 3125 & 131.1 & 0.042 \\ \hline
    scale-2 & 51.49 & 2.82 & 0.001 & 61.8 & 3.214 & 0.002 & 62.69 & 3.576 & 0.002 & 989.6 & 38.93 & 0.011 \\ \hline
    scale-3 & 12.45 & 1.258 & 0.002 & 14.02 & 1.518 & 0.003 & - & - & - & 173.2 & 7.82 & 0.003 \\ \hline
    \multicolumn{12}{c}{} \\ \hline
    \textbf{DES} & \multicolumn{3}{c|}{\textbf{$Q_{poly}$}} & \multicolumn{3}{c|}{\textbf{$Q_{tanh}$}} & \multicolumn{3}{c|}{\textbf{$Q_{opt}$}} & \multicolumn{3}{c|}{\textbf{$Q_{gaus}$}} \\ 
    $\sigma_\epsilon=0.3$, $n_{\rm g}=10$ & 1 & 3 & 5 & 1 & 3 & 5 & 1 & 3 & 5 & 1 & 3 & 5 \\ \hline
    scale-1 & 206.6 & 9.55 & 0.004 & 248.9 & 10.98 & 0.004 & 288.8 & 12.72 & 0.005 & 3593 & 144.5 & 0.043 \\ \hline
    scale-2 & 50.09 & 4.178 & 0.005 & 56.87 & 5.112 & 0.008 & 95.6 & 8.325 & 0.012 & 1047 & 41.67 & 0.012 \\ \hline
    scale-3 & 11.67 & 2.339 & 0.0174 & 11.92 & 2.847 & 0.030 & - & - & - & 169.8 & 9.807 & 0.006 \\ \hline
    \multicolumn{12}{c}{} \\ \hline
    \textbf{CFHTLS} & \multicolumn{3}{c|}{\textbf{$Q_{poly}$}} & \multicolumn{3}{c|}{\textbf{$Q_{tanh}$}} & \multicolumn{3}{c|}{\textbf{$Q_{opt}$}} & \multicolumn{3}{c|}{\textbf{$Q_{gaus}$}} \\
    $\sigma_\epsilon=0.3$, $n_{\rm g}=20$ & 1 & 3 & 5 & 1 & 3 & 5 & 1 & 3 & 5 & 1 & 3 & 5 \\ \hline
    scale-1 & 206.6 & 9.907 & 0.004 & 249.6 & 11.48 & 0.004 & 324 & 14.12 & 0.005 & 3971 & 151.6 & 0.041 \\ \hline
    scale-2 & 49.76 & 4.545 & 0.007 & 55.86 & 5.61 & 0.010 & 104.5 & 9.519 & 0.015 & 1085 & 42.15 & 0.012 \\ \hline
    scale-3 & 11.49 & 2.622 & 0.025 & 11.51 & 3.166 & 0.044 & - & - & - & 169.7 & 10.28 & 0.007 \\ \hline
    \multicolumn{12}{c}{} \\ \hline
    \textbf{Subaru} & \multicolumn{3}{c|}{\textbf{$Q_{poly}$}} & \multicolumn{3}{c|}{\textbf{$Q_{tanh}$}} & \multicolumn{3}{c|}{\textbf{$Q_{opt}$}} & \multicolumn{3}{c|}{\textbf{$Q_{gaus}$}} \\
    $\sigma_\epsilon=0.3$, $n_{\rm g}=30$ & 1 & 3 & 5 & 1 & 3 & 5 & 1 & 3 & 5 & 1 & 3 & 5 \\ \hline \hline
    scale-1 & 198.5 & 16.22 & 0.020 & 219.5 & 21.51 & 0.038 & 603.2 & 42.99 & 0.04045 & 4110 & 160.9 & 0.046 \\ \hline
    scale-2 & 44.84 & 10.82 & 0.117 & 42.97 & 13.32 & 0.237 & 172.8 & 29.81 & 0.1642 & 1070 & 50.42 & 0.021 \\ \hline
    scale-3 & 9.406 & 6.321 & 0.528 & 7.857 & 6.457 & 0.807 & - & - & - & 151.6 & 19.14 & 0.057 \\ \hline
    \multicolumn{12}{c}{} \\ \hline
    \textbf{EUCLID} & \multicolumn{3}{c|}{\textbf{$Q_{poly}$}} & \multicolumn{3}{c|}{\textbf{$Q_{tanh}$}} & \multicolumn{3}{c|}{\textbf{$Q_{opt}$}} & \multicolumn{3}{c|}{\textbf{$Q_{gaus}$}} \\
    $\sigma_\epsilon=0.3$, $n_{\rm g}=40$ & 1 & 3 & 5 & 1 & 3 & 5 & 1 & 3 & 5 & 1 & 3 & 5 \\ \hline \hline
    scale-1 & 194.3 & 20.01 & 0.039 & 206.3 & 27.29 & 0.088 & 730.9 & 58.61 & 0.070 & 4189 & 165.5 & 0.048 \\ \hline
    scale-2 & 42.64 & 14.25 & 0.295 & 38.54 & 16.8 & 0.591 & 197.7 & 40.75 & 0.321 & 1062 & 54.99 & 0.027 \\ \hline
    scale-3 & 8.653 & 7.642 & 1.104 & 6.873 & 7.282 & 1.514 & - & - & - & 143.6 & 24.19 & 0.127 \\ \hline
    \multicolumn{12}{c}{} \\ \hline
    \textbf{LSST} & \multicolumn{3}{c|}{\textbf{$Q_{poly}$}} & \multicolumn{3}{c|}{\textbf{$Q_{tanh}$}} & \multicolumn{3}{c|}{\textbf{$Q_{opt}$}} & \multicolumn{3}{c|}{\textbf{$Q_{gaus}$}} \\
    $\sigma_\epsilon=0.22$, $n_{\rm g}=50$ & 1 & 3 & 5 & 1 & 3 & 5 & 1 & 3 & 5 & 1 & 3 & 5 \\ \hline \hline
    scale-1 & 174.8 & 42.42 & 0.463 & 156.8 & 56.13 & 1.333 & 1206 & 138 & 0.334 & 4169 & 187.7 & 0.070 \\ \hline
    scale-2 & 34.5 & 28.13 & 3.464 & 26.15 & 26.81 & 5.218 & 269.2 & 95.09 & 2.198 & 991.5 & 82 & 0.104 \\ \hline
    scale-3 & 6.403 & 10.32 & 4.964 & 4.519 & 8.139 & 4.889 & - & - & - & 113.5 & 48.96 & 1.688 \\ \hline
    \multicolumn{12}{c}{} \\ \hline
    \textbf{SNAP} & \multicolumn{3}{c|}{\textbf{$Q_{poly}$}} & \multicolumn{3}{c|}{\textbf{$Q_{tanh}$}} & \multicolumn{3}{c|}{\textbf{$Q_{opt}$}} & \multicolumn{3}{c|}{\textbf{$Q_{gaus}$}} \\
    $\sigma_\epsilon=0.3$, $n_{\rm g}=100$ & 1 & 3 & 5 & 1 & 3 & 5 & 1 & 3 & 5 & 1 & 3 & 5 \\ \hline \hline
    scale-1 & 172.6 & 45.42 & 0.5824 & 152.5 & 59.33 & 1.664 & 1322 & 148.6 & 0.3481 & 4287 & 190.2 & 0.069 \\ \hline
    scale-2 & 33.73 & 29.39 & 4.133 & 25.22 & 27.43 & 6.009 & 281.6 & 102.2 & 2.494 & 991.3 & 85.32 & 0.117 \\ \hline
    scale-3 & 6.218 & 10.41 & 5.403 & 4.355 & 8.1 & 5.19 & - & - & - & 110.8 & 51.68 & 2.083 \\ \hline
  \end{tabular}
\end{table*}

\end{document}